\documentclass[aps,prb,twocolumn,superscriptaddress,amsmath]{revtex4} 

\usepackage{amssymb,graphicx,bm}

\usepackage{amsfonts}
\usepackage{bbold}
\usepackage{color}

\newif\ifpdf
\ifx\pdfoutput\undefined
\pdffalse
\else
\pdfoutput=1
\pdftrue
\fi
\ifpdf
 \usepackage{epstopdf}
\else
\fi

\begin{document}

\title{Y-junction connecting Luttinger liquids: fixed point structure and
conductances}
\author{D.N. Aristov}
\affiliation{Petersburg Nuclear Physics Institute, Gatchina 188300, Russia}
\affiliation{ Institute for Nanotechnology, Karlsruhe Institute of Technology,
76021 Karlsruhe, Germany } 
\affiliation{Department of Physics, St.Petersburg State University, 
Ulianovskaya 1, Petrodvorets, St.Petersburg 198504, Russia}  
\author{P. W\"olfle}
\affiliation{ Institute for Nanotechnology, Karlsruhe Institute of Technology,
76021 Karlsruhe, Germany }
\affiliation{Institute for Condensed Matter Theory, and
Center for Functional Nanostructures, Karlsruhe Institute of Technology,
76128 Karlsruhe, Germany}
\date{\today}

\begin{abstract}
We study the transport properties of three Luttinger liquid wires (with
possibly different interaction strength), connected through a Y-junction,
within the scattering state formalism. We first formulate the problem in
current algebra language and focus on the case of a symmetric set-up, for
which the scattering matrix and the matrix of conductances is parametrized
by two variables. For these we derive coupled RG equations, first in a
ladder summation up to infinite order in the interaction. The fixed point
structure and the implicit solution of these equations is presented. It is
shown that higher order terms beyond the ladder approximation do not change
the scaling behavior near the fixed points. For sufficiently strong
attractive interaction a new fixed point with unusual properties is found.
\end{abstract}

\maketitle



\section{\label{sec:intro} Introduction}


Electron transport in strictly one-dimensional quantum wires is governed by
the Coulomb interaction between electrons. This is spectacularly
demonstrated by the fact that within the Tomonaga-Luttinger liquid (TLL)
model in the limit of temperature $T\rightarrow 0$ the conductance of a
quantum wire with finite barrier tends to zero, provided the interaction is
repulsive and assumes the maximum value $G=1$ (in units of the conductance
quantum $G_{0}=e^{2}/h$ ) for attractive interaction. The latter behavior
appears independent of the strength of the scattering at the barrier, and
can be traced back to the formation of Friedel oscillations of the charge
density around the barrier, leading to an infinitely extended effective
barrier potential in that limit. 

TLL behavior has recently been studied
experimentally in carbon nano-tubes. \cite{Bockrath1999,Yao1999} 
In a future nanoelectronics
constructed out of quantum wires, junctions of three or more wires will
necessarily be involved, requiring a knowledge of the fundamental
transport behavior of electrons in such structures. It is known that such
systems exhibit rather rich TLL effects which have been the subject of a
number of recent papers. 
\cite{Nayak1999,Safi2001,Yi2002, Lal2002,  Chen2002,Moore2002,
Chamon2003,Pham2003,Egger2003, 
Kim2004,Rao2004,Kazymyrenko2005,Enss2005,Barnabe2005,Barnabe2005a}
Much of the work in this
field has used the bosonization method, which gives rise to the problem of how to preserve the fermionic character of charge carriers. When the
number of wires meeting at a junction exceeds two, the Klein factors, which
give Fermi statistics to the bosonized operators of different wires, are more difficult to handle. \cite{Nayak1999}
Oshikawa et al. recently introduced \cite{Oshikawa2006} a new method to study this problem,
mapping it into the dissipative Hofstadter model (DHM), which describes a
single particle moving in a uniform magnetic field and a periodic potential
in two dimensions and coupled to a bath of harmonic oscillators. When the
three quantum wires enclose a magnetic flux, the mapping to the DHM also
allows to identify a new low energy chiral fixed point with an asymmetric
flow of current that is highly sensitive to the sign of the flux. The paper by \cite{Oshikawa2006} is giving a systematic and thorough treatment of the three lead junction problem, and we shall refer to it later, when we compare our results with theirs. It should be noted, however, that in spite of their success, the above mentioned methods do not in general allow to determine the fixed point (FP) structure of the theory in an unbiased way, as they require a certain knowledge of the existence of FPs as a starting point.

In this paper we follow a different route: we describe the transport
properties in fermionic language, thus avoiding the problem of Klein
factors and the necessity to assume the existence of FPs. As shown in our previous work, one may recover the known results on
transport through a TLL wire with barrier and obtain new results not
accessible by other methods within the fermionic formulation. \cite{Aristov2009} Moreover,
as this method is formulated in a scattering wave picture, the connection of
the interacting wire to non-interacting reservoirs is naturally included.
While this formulation has previously been regarded as being restricted to
weak coupling \cite{Yue1994,Das2004,Lal2002} we have shown in an earlier work that it may
be extended to the strong coupling regime in a controlled way by using an
infinite resummation of perturbation theory (ladder summation). The result
obtained in this approximation is universal, i.e. does not depend on the
ultraviolet cutoff chosen. 

We restrict ourselves to a symmetric time-reversal invariant setup,
which is characterized by two independent components of the conductance tensor. 
This setup in case of one wire without interaction is relevant to the problem of tunneling into a Luttinger liquid and was studied   in \cite{Aristov2010} for weak interaction.
There it has been found that the asymmetric fixed point A (perfectly conducting wire and a vanishing tunneling amplitude into the tip, see below) becomes unstable in the case of weak repulsive interaction once the description of the Y-junction is not reduced to the simple tunneling model. A further remarkable result of that work is the finding of a nonmonotonic behavior of the conductance (e.g. as a function of temperature), in certain cases. We extend that work here to arbitrary interaction and find similar behavior at strong interaction. 

The resulting coupled RG equations will first be analyzed with respect to their fixed point
structure. We find four fixed points, labelled N,A,M,Q, where at N (Neumann) the three 
wires are totally separated, at A (asymmetric) the third wire is separated, while the main 
wire is perfectly conducting, at M (mystery) and Q (quaint) the conductances take an 
intermediate value, depending on the interaction strength. All fixed points are located on 
the boundary curve marking the allowed area in the plane of the two conductances.  For 
repulsive interaction only N is stable, A being a saddle point. For attractive interaction N is 
unstable and M is stable. For stronger attractive interaction, beyond a threshold value A 
becomes an additional stable fixed point. The conductances are calculated analytically for 
several special cases of interest. We present and discuss the power law exponents 
appearing in the various regimes. 

In order to check the reliability of the ladder summation result, we
calculate all scale dependent terms (in the limit of $T=0$ ) up to third order in the interaction. We
find several terms additional to the ladder terms. These terms are
nonuniversal and subleading, in the sense that they disappear for the case of repulsive interaction, when one of
the stable fixed points is approached. For attractive interaction, the properties of the new fixed point M appear to be affected by the three-loop terms beyond the ladder approximation.

The paper is organized as follows. In section \ref{sec:Setup} we define the model of a Y-
junction considered here. We introduce the currents in the SU(3)-representation and
define the Hamiltonian in the chiral representation. Section \ref{sec:SmatCond} is devoted 
to an analysis of the scattering matrix and the derivation of the conductance
tensor in terms of the S-matrix components. We consider a simple tunneling case and the 
totally symmetric case, both characterized by only a single free parameter in the S-matrix 
and the general Y-junction case with two parameters and therefore two independent
conductance components. In section \ref{sec:RGeq} we show how the RG equation for the
simple tunneling case is extracted from perturbation theory in the ladder
approximation, and discuss the ensuing conductance. The RG equation for the general Y-junction case is presented in section \ref{sec:GenSym}.
We further present the
results on the nonuniversal terms in third order perturbation theory in section \ref{sec:perturb1}.
Finally, the RG phase diagram is presented and discussed in section \ref{sec:discussion}.

\section{\label{sec:Setup} Model of a Y-junction}

We consider the setup of a Y-junction: a quantum wire of finite length $2L$
connected to noninteracting leads, and a tip with attached wire forming a
contact (junction) in the center of the wire, which for simplicity is
assumed to have the same length $L$, 
also connected to a noninteracting lead. We choose the origin of the coordinate
system at the junction and denote the two halves of the main wire by indices
1 and 2 and the tip by 3. We consider spinless fermions. In the following we will refer to the so defined symmetric (leads $1,2$) set up as a Y-junction. A particularly simple version to be considered below is a simple tunneling junction (no tunneling barriers, only on site tunneling). Below we will also briefly consider a totally symmetric junction (with respect to interchanging any pair of wires).

The electrons are assumed to interact via a short-ranged interaction of arbitrary strength and sign between
incoming and outgoing fermions. In the present case, instead of defining
right- and left-movers, it is more appropriate to speak of incoming and
outgoing waves, with respect to the junction, or ``origin''.

The short-range interaction is characterized by the amplitude $g$ in the
main wire ( 1 and 2) and a different amplitude $g_3$ in the third wire. This
interaction takes place at $x < L$, measuring from the origin. In order to 
spatially separate effects of the potential scattering at the origin and the interaction, we 
also assume that the interaction takes place at $x>a_{0}$, whereas the 
potential scattering happens at $x<a_{0}$. The scale $a_{0}$ appears below as 
short-range cutoff of the theory.   

The Hamiltonian of the system is
\begin{eqnarray}
\mathcal{H} &=& \int_{-\infty}^0 dx \sum_{j=1}^3 H_j \\
H_j &=& v_F \psi^\dagger_{j,in} i \nabla \psi_{j,in} - v_F
\psi^\dagger_{j,out} i \nabla \psi_{j,out} \notag \\
&& + 2\pi v_F g_j \psi^\dagger_{j,in} \psi_{j,in} \psi^\dagger_{j,out}
\psi_{j,out}  \label{iniHam}
\end{eqnarray}
where we did not write explicitly the range function of the interaction. In
the chosen parametrization the amplitudes $g_j$ are dimensionless and enter
the subsequent expressions in the most convenient form.

\begin{figure}[tbp]
\includegraphics*[width=0.75\columnwidth]{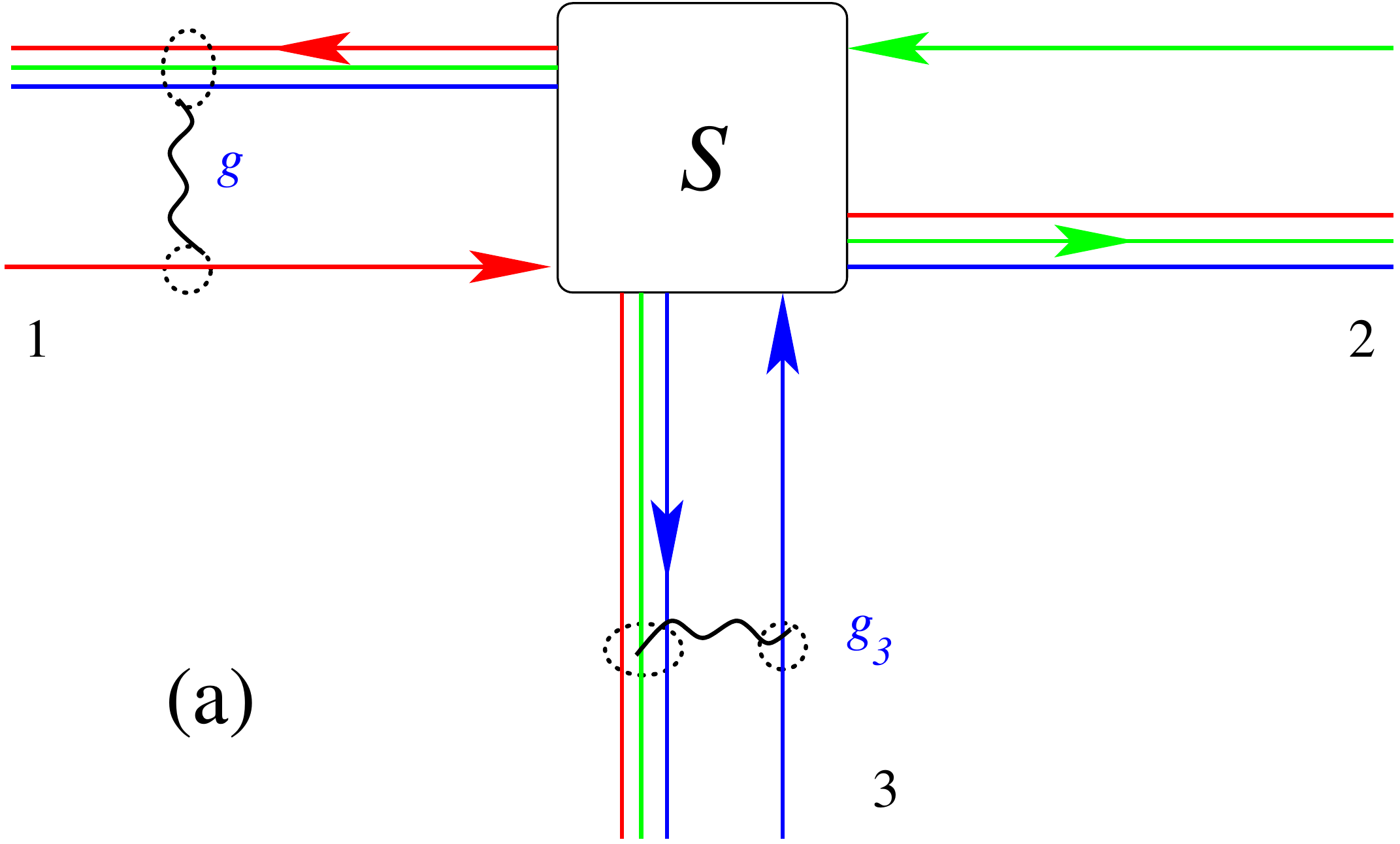} %
\includegraphics*[width=0.75\columnwidth]{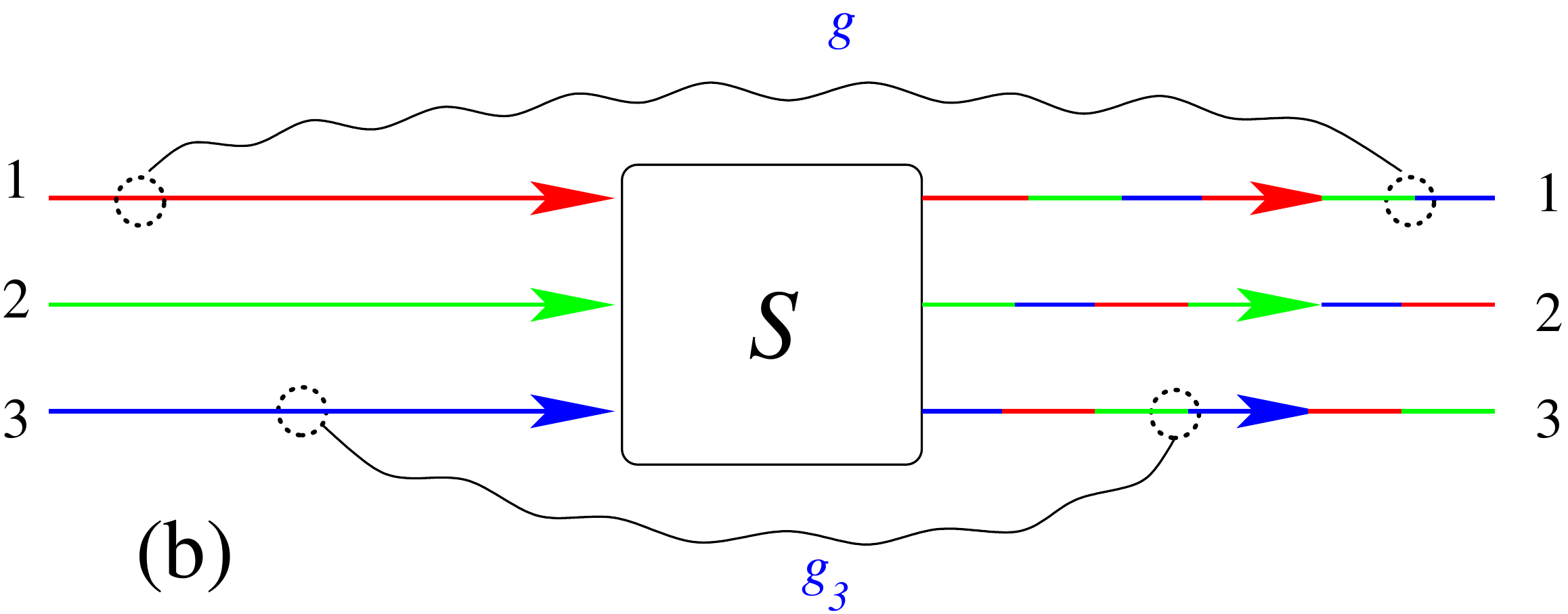}
\caption{(Color online) (a) The geometry of a $Y$-junction is shown together
with currents of incoming and outgoing fermions, the type of fermions in the
scattered states representation is indicated by its color. The local
short-range interaction between the fermions is shown by a wavy line. (b)
The equivalent representation in the chiral fermion basis after the
unfolding procedure; the initially local interaction becomes non-local, as
described by Eq.\ (\protect\ref{eq:Ham2}). }
\label{fig:arrows}
\end{figure}

We unfold the setup in the usual way, \cite{Affleck1991,Fabrizio1995} by
putting the incoming fermions on the negative $x$-axis and the fermions,
which have passed through the junction, at positive $|x|$. As a result, we
have a three-fold multiplet of right-going fermions with non-local
interaction. This procedure is depicted in Fig. \ref{fig:arrows}

The boundary condition at the origin is described by the scattering S-matrix
as follows. For elastic scattering by the central dot, the outgoing fermions
at the origin are connected to the incoming ones by the relation $%
\psi^\dagger_{k,out} = S_{km}^{\ast} \psi^\dagger_{m,in}$. We choose $S$ in the
symmetric way as described at length in the next subsection.

In the scattered states representation, the right- and left- going fermionic
densities acquire the form 
\begin{equation}
\begin{aligned} \psi_{j,in} (x) &= \psi_{j} (x), \quad \psi_{j,out} (x) =
(S.\psi)_{j} (-x) \\ \psi^\dagger_{j,in} (x) \psi_{j,in} (x) &=
\rho_j(x), \quad \psi^\dagger_{j,out} (x) \psi_{j,out} (x) =\tilde
\rho_j(-x) \end{aligned}
\label{def:scatstates}
\end{equation}
here and below we use the notation $\tilde A = S^\dagger . A. S$.

Let us explain the meaning of these quantities. We introduce a multiplet of
incoming fermions $\Psi = (\psi_1, \psi_2, \psi_3)$. The incoming density $%
\rho_j =\Psi^\dagger .\hat \rho_j .\Psi$ is given by a diagonal matrix, i.e.

\begin{equation}
\begin{aligned}
\hat{\rho}_{1} &=\begin{pmatrix} 1 & 0 &0\\ 0 & 0 &0\\ 0 & 0 &0
\end{pmatrix}=\frac{1}{2}\left( \sqrt{\frac{2}{3}}\lambda _{0}+\frac{1}{%
\sqrt{3}}\lambda _{8}+\lambda _{3}\right)  \\
\hat{\rho}_{2} &=\begin{pmatrix} 0 & 0 &0\\ 0 & 1 &0\\ 0 & 0 &0
\end{pmatrix}=\frac{1}{2}\left( \sqrt{\frac{2}{3}}\lambda _{0}+\frac{1}{%
\sqrt{3}}\lambda _{8}-\lambda _{3}\right)  \\
\hat{\rho}_{3} &=\begin{pmatrix} 0 & 0 &0\\ 0 & 0 &0\\ 0 & 0 &1
\end{pmatrix}=\frac{1}{\sqrt{6}}\lambda _{0}-\frac{1}{\sqrt{3}}\lambda _{8}
\end{aligned}
\label{rhos2lambdas}
\end{equation}
Here and
below $\lambda _{j}$, with $j=1,\ldots 8$ 
are the traceless Gell-Mann matrices,  discussed in the Appendix \ref{sec:GellMann}.
 In addition to these, we use also the matrix $\lambda
_{0}=\sqrt{\tfrac{2}{3}}\,\mathbb{1}$, which is proportional to the unit
matrix, $\mathbb{1}$. 

The outgoing densities are given by $\tilde \rho_j =\delta_{jk} S^\ast_{kl}
S_{km} \psi^\dagger _l \psi_m = \Psi^\dagger . S^\dagger . \hat \rho_j .S.
\Psi$, so that in the matrix representation $\hat{\tilde \rho}_j = S^\dagger
. \hat \rho_j . S $. We will mostly omit the hat sign over $\rho$ below.

In terms of the above densities the interaction terms may now be written.
\begin{equation}
\begin{aligned} & \rho_1(-x) \tilde \rho_1(x) + \rho_2 (-x) \tilde \rho_2
(x) \\ & = \tfrac12 \Psi^\dagger(-x) \left(\sqrt{\tfrac23} \lambda_0 +
\tfrac1{\sqrt{3}} \lambda_8 \right)\Psi(-x) \\
& \times \Psi^\dagger(x)
\left(\sqrt{\tfrac23} \lambda_0 + \tfrac1{\sqrt{3}} \tilde \lambda_8
\right)\Psi(x) \\& +\tfrac12 \Psi^\dagger(-x) \lambda_3 \Psi(-x)
\Psi^\dagger(x) \tilde \lambda_3 \Psi(x) \end{aligned}
\end{equation}
where $\tilde \lambda_j=S^\dagger. \lambda_j . S $, or simply

\begin{equation}
\begin{aligned} \rho_1\tilde \rho_1 + \rho_2 \tilde \rho_2 & =
\left[\frac1{\sqrt{3}} \lambda_0 + \frac1{\sqrt{6}} \lambda_8 \right]
\left[\frac1{\sqrt{3}} \lambda_0 + \frac1{\sqrt{6}} \tilde \lambda_8 \right]
\\ &
+\frac12 \lambda_3 \tilde \lambda_3  \equiv \rho_+ \tilde \rho_+ +
\rho_- \tilde \rho_- \end{aligned}
\label{def:rho+-}
\end{equation}
\begin{equation}
\begin{aligned} & \rho_3\tilde \rho_3 = \left( \frac1{\sqrt{6}} \lambda_0
-\frac1{\sqrt{3}} \lambda_8 \right) \left( \frac1{\sqrt{6}} \lambda_0
-\frac1{\sqrt{3}} \tilde \lambda_8 \right) \end{aligned}
\end{equation}
Here we defined $\rho _{+}=\frac{1}{\sqrt{3}}\lambda _{0}+\frac{1}{\sqrt{6}}%
\lambda _{8}$, $\rho _{3}=\frac{1}{\sqrt{6}}\lambda _{0}-\frac{1}{\sqrt{3}}%
\lambda _{8}$, and $\rho _{-}=\frac{1}{\sqrt{2}}\lambda _{3}$.

In terms of these quantities, the Hamiltonian takes the following form (from
now on we set $v_F=1$) 
\begin{equation}
\begin{aligned} {\cal H} &= \int_{-\infty}^\infty dx\, 
\Big[ \sum_{j=1,\ldots 3} \psi^\dagger_{j} i \nabla \psi_{j} 
\\ & + 2\pi \Theta(a<x<L)
\left( g\rho_+ \tilde \rho_+ + g \rho_- \tilde \rho_- + g_3 \rho_3 \tilde
\rho_3 \right) \Big ] 
\end{aligned}  \label{eq:Ham2}
\end{equation}
where $\rho_\pm , \rho_3$ refer to $-x$ and $\tilde \rho_\pm , \tilde \rho_3 
$ refer to $x$, and $\Theta(a_{0}<x<L)$ is equal to $1$ within the interval
specified and zero elsewhere.

For simplicity we do not consider here the so-called $g_{4}$ part of the
fermionic interaction, i.e. the terms 
$  \pi v_F \bar g_j [ (\psi^\dagger_{j,in} \psi_{j,in})^{2}  +  (\psi^\dagger_{j,out}
\psi_{j,out}  ) ^{2} ] $. It is known that the $g_{4}$-interaction 
can be absorbed into the redefinition of the group velocity inside the interacting region, 
$\tilde{v}_{Fj}=v_{F}(1+\bar g_{j} ) $. For finite $L$ one can show that $g_{4}$
does not lead to a renormalization of d.c.\ conductance, which is the
quantity of our interest below. The effect of $g_{4}$ on the a.c. conductance
can be analyzed, e.g., following the guidelines in \cite{Safi1995}.

\section{\label{sec:SmatCond}S-matrix and conductances}

\subsection{\label{sec:introSmat}S-matrix}

The most general S-matrix is defined as follows 
\begin{equation}
S = \begin{pmatrix} r_{1}, & t_{12}, & t_{13}\\ t_{21} , & r_{2}, & t_{23}
\\ t_{31} & t_{32} & r_{3} \end{pmatrix}
\end{equation}
where $r_{j}$ is the reflection amplitude for wire $j$, and $t_{jk}$ is the
transmission amplitude between wires $j$ and $k$. The matrix $S$ is unitary $%
S^\dagger S = 1$, which allows its parameterization via the exponential, $S =
\exp \left( i \sum_{j=0}^8 \theta_j \lambda_j\right)$. Obviously, there is a
redundancy in the description by the nine real-valued parameters $\theta_{j}$ in the
exponent, as only the densities and not the fermion amplitudes enter the
observables.

The number of physically relevant parameters for the description of the $S$%
-matrix may be determined as follows. One may fix the relative $U(1)$ phase
between the ingoing and outgoing electrons in each wire, by demanding that
the reflection coefficients are real valued. This excludes the $\lambda_3$
and $\lambda_8$ components. It turns out that it is more convenient to keep 
the $\lambda_8$ component, at the cost of introducing some redundancy, 
see below. In addition the overall phase described by the $%
\lambda_0$ term may be set to zero.

A further reduction in the number of independent parameters may be derived
from the following consideration. In the limit of three almost detached
wires, we have $S\simeq \mathbb{1} + i \sum_{j=0}^8 \theta_j \lambda_j $,
and $\theta_j \ll 1$. This limiting case elucidates the meaning of the $%
\theta_j $ as the tunneling amplitudes between the corresponding wires. For
example, the infinitesimal hopping between wires 1 and 3 of the form $%
(t_{13} \psi_{3}^{\dagger} \psi_{1} +h.c.) \vert_{x=0} $ leads to $%
\theta_{4} = Re(t_{13})/v_{F}$, $\theta_{5} = Im(t_{13})/v_{F}$ with $v_{F}$
the Fermi velocity, cf.\ \cite{Oshikawa2006}. It should be noted that this
correspondence between the Hamiltonian and the $S$-matrix holds only in the
limiting cases, and starts to depend on the definition of the regularization
procedure in higher orders of $t_{kl}$, i.e.\ beyond the Born approximation. 
\cite{Aristov2009}

By using the above correspondence of the (small) $\theta_{j}$ and the
tunneling amplitudes it becomes clear that without loss of generality one
may require real-valued hopping amplitudes $t_{13}$ and $t_{23}$, i.e.
coinciding phases for hopping from the third wire into wires $1$ or $2$ (in the absence of magnetic flux).
This makes $\theta_5=\theta_7=0$. After that the phase of the complex valued
amplitude $t_{12}$ between the wires 1 and 2 is fixed. In the absence of
magnetic fields it must be zero \cite{Oshikawa2006} anyway, which further
reduces the number of components, since it requires $\theta_2=0$.

Thus in the time-reversal invariant situation, the S-matrix can be
parametrized by three angles, $\theta _{1},\theta _{4},\theta _{6}$. In the
presence of a magnetic field a fourth component, $\theta _{2}$,
appears. We do not consider this case here. 

In this paper we concentrate on the case of tunneling into the center of a Luttinger liquid wire. This amounts to full symmetry between the wires 1 and 2, and it reduces the
number of independent parameters further to only two, $\theta _{1}$ and $%
\theta _{4}=\theta _{6}$.

As mentioned above, we will in addition keep the angle $\theta _{8}$. 
The reason for the introduction of the $\lambda _{8}$ component is two-fold.
First, the explicit analytic expressions for $S$ below are somewhat
simplified. Second, even if we choose to start without $\lambda _{8}$
component, it will be generated during the renormalization process, as we
show below in Sec.\ \ref{sec:GenSym}. This means that the three angles 
$\theta _{j}, j=1,4,8$, are not independent.  Below we will identify
proper combinations of these variables forming a minimal set of two
independent variables, in terms of which all the other quantities may be
expressed. We then parametrize the three angles in terms of a set of 
new angles $\tilde{\theta},\phi ,\tilde{\psi}$ as follows 

\begin{equation}
\begin{aligned} \sum_{j=1}^8 \theta_j \lambda_j &= \tilde \theta \cos \phi
(\lambda_{4}+ \lambda_{6})/\sqrt{2} \\ & + \tfrac12 (\tilde\theta\sin \phi +
\tilde \psi) \lambda_{1} 
+ \tfrac1{2\sqrt{3}} (3 \tilde\theta \sin \phi
- \tilde\psi) \lambda_{8} \end{aligned}  \label{SmatGen1}
\end{equation}

Explicitly we have the representation of the $S$-matrix in terms of three
angles $\tilde{\theta},\phi ,\tilde{\psi}$  
\begin{equation}
S=e^{i\tilde{\psi}/3}\begin{pmatrix} r_{1} & t_{1} & t _{2}\\ t_{1} & r_{1}
& t _{2}\\ t_{2} & t_{2}& r_{2}\end{pmatrix}
\end{equation}%
with 
\begin{equation}
\begin{aligned} r_{1} &= \tfrac 12 (e^{-i \tilde\psi} + \cos \tilde\theta +
i \sin \tilde\theta \sin \phi) \\ t_{1} &= \tfrac 12 (-e^{-i \tilde\psi} +
\cos \tilde\theta + i \sin \tilde\theta \sin \phi) \\ t_{2} &= \tfrac
i{\sqrt{2}} \sin \tilde\theta \cos \phi \\ r_{2} &= \cos \tilde\theta - i
\sin \tilde\theta \sin \phi \end{aligned}  \label{genSparam1}
\end{equation}%

To identify one of the variables as redundant, we use the transformation $(%
\tilde{\theta},\phi ,\tilde{\psi})\rightarrow (\theta ,\psi ,\gamma )$, with

\begin{equation}
\begin{aligned} \tan \gamma &= \sin \phi \tan \tilde\theta , 
\quad  \psi  =\tilde \psi + \gamma , \\ \cos\theta & = \frac{\cos \tilde\theta}{\cos \gamma} .
\end{aligned}
\end{equation}

After some calculation we find the components of the $S$-matrix 
(up to an overall phase factor $e^{i(\psi -\gamma )/3}$) as

\begin{equation}
\begin{aligned} r_{1} &=\tfrac 12 (e^{-i \psi} + \cos \theta ) e^{i\gamma}
\\ t_{1} &= \tfrac 12 (-e^{-i \psi} + \cos \theta ) e^{i\gamma} \\ t_{2} &=
\tfrac i{\sqrt{2}} \sin \theta \\ r_{2} &= \cos \theta e^{-i\gamma}
\label{genSparam2} \end{aligned}
\end{equation}

We show in the next subsection that the new third variable $\gamma $ does 
not appear in any of the
components of the conductance. In addition, we will see later 
in Sec.\ \ref{sec:GenSym} that $\gamma $
does not take part in the renormalization process. 

It will turn out to be useful to consider the most elementary case of tunneling from the third wire into the main wire (no next neighbor tunneling or tunneling barrier) separately. In that case the S-matrix is characterized by a single angle, $\theta =\sqrt{2} \theta_{4}=\sqrt{2} \theta_{6}$,
whereas $\theta_{1}=0$. In terms of the above variables we have $(\tilde{\theta},\phi ,\tilde{\psi})=(\theta,0,0)$  or else $(\theta ,\psi ,\gamma )=(\theta ,0,0)$.

In the fully symmetric case of a junction of three identical wires (identical interaction strength and reflection coefficients) we have only one independent angle characterizing the S-matrix, $\theta=\theta_{1}=\theta_{4}=\theta_{6}$. The connection to the above $(\tilde{\theta},\phi ,\tilde{\psi})$ ,  $(\theta ,\psi ,\gamma ) $ is not very transparent, and up to an overall phase factor  we have $r_{1}=r_{2}= (2+e^{3i\theta})/3$,  $t_{1}=t_{2}= (-1+e^{3i\theta})/3$. 

\subsection{Conductances}

\label{sec:Conduc}

The observables we concentrate on in this work are the linear conductances.
We first discuss the question of the number of independent linear
conductances. The matrix of conductances, $G_{jk}$, is defined through the
relation connecting the current $I_{j}$ in a given wire with the electric
potentials $V_{k}$ in all the leads as $I_{j}=\sum_{k}G_{jk}V_{k}$. The
current in the $j$th wire is given by $I_{j}(x)=ev_{F}(\langle \rho
_{j,in}(x)\rangle -\langle \rho _{j,out}(-x)\rangle )$. The electric
potentials $V_{k}$ give rise to the following source term in the Hamiltonian, $%
H_{V}=e\int_{-\infty }^{0}dx\sum_{k}V_{k}(\rho _{k,in}(x)+\rho _{k,out}(-x))$%
. In linear response theory \cite{Aristov2009} the currents are given by $%
I_{j}(x)=\int_{-\infty }^{0}dy(\langle \rho _{j}(x)\rho _{k}(y)\rangle
-\langle \tilde{\rho}_{j}(-x)\rho _{k}(y)\rangle )V_{k}$ In the static limit
and in the absence of interaction the response functions may be evaluated to
give 
\begin{equation}
G_{jk}=\delta _{jk}-Tr(\hat{\tilde{\rho}}_{j}\hat{\rho}_{k})=\delta
_{jk}-|S_{jk}|^{2}
\label{GtoS}
\end{equation}%
One easily verifies that the charge is conserved, $\sum_{j}G_{jk}=0$, and
that applying equal potentials to all wires produces no current, $%
\sum_{k}G_{jk}=0$.

In view of these conservation laws, it is more instructive to discuss the
current response to certain  bias voltages. Let us define $%
(I_{a},I_{b},I_{0})=\widehat{G}.(V_{a},V_{b},V_{0})$, with 
\begin{equation}
\begin{aligned} V_a &= (V_1 - V_2), \quad I_a  = (I_1 - I_2)/2, \\ V_b &=
(V_1 + V_2 - 2V_3)/2 \quad I_b = (I_1 + I_2 - 2I_3)/3 , \\ V_0 & = (V_1 +
V_2+V_{3})/3 , \quad I_0 = (I_1 + I_2+ I_3) /3, \end{aligned}
\end{equation}%
In this notation, we seek the retarded response of the current $I_{a}(x)=%
\tfrac{1}{2}ev_{F}(\langle \psi ^{\dagger }\lambda _{3}\psi (x)\rangle
-\langle \psi ^{\dagger }\tilde{\lambda}_{3}\psi (-x)\rangle )$, to the
source term , e.g., $H_{V,a}=\frac{e}{2}\int_{-\infty }^{0}dxV_{a}\psi ^{\dagger
}\lambda _{3}\psi (x)$. We can write these combinations
symbolically in the static limit as 
\begin{equation}
\begin{aligned} H_{V} &=  \tfrac12 V_a \lambda_{3}+ \tfrac1{\sqrt{3}} V_b
\lambda_{8} +  \tfrac1{\sqrt{6}} V_0 \lambda_{0} , \\ I_a & = \tfrac12
(\lambda_{3} - \tilde \lambda_{3}), \quad I_b = \tfrac1{\sqrt{3}}
(\lambda_{8} - \tilde \lambda_{8}), \\ I_0 &=\tfrac1{\sqrt{6}} (\lambda_{0}
- \tilde \lambda _{0} )\to 0, \end{aligned}
\end{equation}%
As expected, $I_{0}=0$; in addition we may choose the zero of the electric
potentials such that $V_{0}=0$. It can be shown \cite{Aristov2009} that the
conductance in the d.c.\ limit is proportional to the trace of the product of
current and the source vertices. It follows then that the line and the row,
corresponding to $I_{0}$ and $V_{0}$ are identically zero, and we omit them
for clarity below. The remaining four components are non-zero and we obtain
the conductance matrix in the general form 
\begin{equation}
\begin{aligned}
\hat{G} & =\begin{pmatrix} G_{aa} & G_{ab} \\ G_{ba} & G_{bb} \\ \end{pmatrix}
\\&  =
\begin{bmatrix} \tfrac12 \left(1 - \tfrac12 Tr(\tilde\lambda_{3} \lambda_{3}) \right), &
- \tfrac 1{2\sqrt{3}} Tr(\tilde\lambda_{3} \lambda_{8}) \\ - \tfrac
1{2\sqrt{3}} Tr(\tilde\lambda_{8} \lambda_{3}) , & \tfrac23 \left( 1 - \tfrac12
Tr(\tilde\lambda_{8} \lambda_{8}) \right) \\ \end{bmatrix} 
\end{aligned}
\label{defGprime}
\end{equation}

\subsubsection{Conductance of Y-junction}

For the particular choice of $S$ given in (\ref{genSparam2}) above we have 
\begin{equation}
\begin{aligned} \hat G & = \begin{pmatrix}  
|t_{1}^{2}| + \tfrac 12 |t_{2}^{2}|, &0 \\ 0, & 2|t_{2}^{2}|   \end{pmatrix} \\ & =
\begin{pmatrix} \tfrac12 [1 - \cos \theta \cos \psi ], &0 \\ 0, &  \sin^2 \theta
\end{pmatrix}  \end{aligned} 
\label{Gmatrixhat}
\end{equation}

We see that the new third variable $\gamma $ does not appear in any of the
components of the conductance. In addition, we show below that $\gamma $
does not take part in the renormalization process. Hence, in line with the
above argument, the general symmetric case is completely determined by two
independent parameters, which can be ultimately chosen as the two non-zero
components of the conductance matrix, $G_{aa}\equiv G_{a},$ $G_{bb}\equiv
G_{b}$, Eq.\ (\ref{Gmatrixhat}). We will present two coupled RG equations for
the two conductances $G_{a},G_{b}$ in Sec.~\ref{sec:GenSym} below.

It follows from the parametrization (\ref{Gmatrixhat}) of the conductances that the physically accessible area in the two-dimensional space of conductances is not simply given by the unit square, but is defined by 

\begin{equation}
0\leq G_{b}\leq 1-4\left(G_{a}-\tfrac{1}{2}\right)^{2}
\label{bound}
\end{equation}

We will see below that the boundary of the physical regime in $G_{a},G_{b}$-space defined by (\ref{bound}) plays a special role in that all fixed points of the problem are located on the boundary. We will use this fact to our advantage when we consider the simplified structure of the RG-flow along the boundary curve.

In the most general case, the matrix of conductances has four independent
components, in accordance with the analysis of the number of relevant
parameters characterizing the general $S$-matrix given in  \cite{Aristov2011}.

It is worth to note the following property of the boundary (\ref{bound}). It was proven in \cite{Bengtsson2005} that the boundary of the region of allowed conductances, defined by Eq.\ (\ref{GtoS}), corresponds to the matrices $S$ which can be made entirely real by certain ``rephasing'', i.e.\ multiplication of rows and columns of $S$ by phase factors. It was also proven that one can recover $S$ from $G$ up to this ``rephasing''. \cite{Bengtsson2005}
In our particular case (\ref{genSparam2}) the ``rephasing'', leading to the real-valued last line and last column, is achieved by $diag[i e^{-i\gamma},i e^{-i\gamma},1] . 
S. diag[i,i, e^{i\gamma}]$. It is then clear, that the boundary corresponds either to $\psi=0,\pi$, or $\theta=0,\pi$, which is indeed the case, as shown below (see also \cite{Aristov2011}). 

\subsubsection{Conductance in the simple tunneling case}

Let us now consider the special case when the S-matrix is characterized by 
only a single parameter, $\theta=\theta_4=\theta_6$. This corresponds to the simplest 
model of tunneling from the tip into the wire.  Then we have 

\begin{equation}\begin{aligned} 
\tilde \lambda_3 & = \cos\theta \lambda_3 -
\tfrac1{\sqrt{2}} \sin\theta( \lambda_4 - \lambda_6 )
\\ \tilde \lambda_8 &
= -\tfrac{\sqrt{3}}2 \sin^2 \theta \lambda_1 - \sqrt{\tfrac38} \sin2\theta(
\lambda_4 + \lambda_6 )
\\&  + \frac{1+3\cos2\theta}4 \lambda_8
\end{aligned} \label{rotL38} \end{equation}%

and the reduced conductance tensor follows as
\begin{equation}
\hat{G}=\begin{pmatrix} \sin^2 \tfrac {\theta}2 &0 \\ 0, & \sin^2 \theta \\
\end{pmatrix}
\label{Gmatrix2}
\end{equation}

We note that the conductances satisfy the relation $G_{b}=1-4(G_{a}-\frac{1}{2})^{2}$ , implying that the simple tunneling case traces the boundary of the physically allowed region  in $G_{a},G_{b}$-space, see (\ref{bound}).  As we will see below this case is more than a specialization to a very simple physical model. It actually already contains the information on the fixed point structure of the RG flow of the general model. Its advantage is that the fixed points and even the conductances may be calculated analytically. The stability of the fixed points cannot be decided in this restricted model, as runaway flow away from the boundary may occur (see Sec.\ \ref{sec:GenSym} below).

\subsubsection{Conductance in the fully symmetric case}

For the fully symmetric case with $t_{1}=t_{2}=\tfrac13 (e^{3i\theta}-1)$ we have from Eq.\ (\ref{Gmatrixhat}) 
\begin{equation}
\hat{G}=\begin{pmatrix} \tfrac23 \sin^2 \tfrac {3\theta}2 &0 \\ 
0, & \tfrac89 \sin^2 \tfrac {3\theta}2 \\
\end{pmatrix}
\end{equation}
which shows that the maximum transparency of the fully symmetric junction is reached at $\theta = \pi/3$ and corresponds to $G_{a}=2/3$, $G_{b}=8/9$. These values are smaller than the maximum (unity) values of the individual conductances, which according to  (\ref{bound}) are attained at the points $G_{a}=1/2$, $G_{b}=1$ and $G_{a}=1$, $G_{b}=0$ in the $G_{a}-G_{b}$ - plane. 

\section{\label{sec:RGeq}RG equation: universal contributions}

\subsection{Simple tunneling case}

\subsubsection{\label{sec:RGtun1st}Lowest order}

The renormalization of the $S$-matrix due to interaction can be understood
by considering the simplest diagrams in perturbation theory. To illustrate
our approach, we consider first the special case of the $S$-matrix, 
$(\tilde{\theta},\phi ,\tilde{\psi})=(\theta,0,0)$ in Eq.\ (\ref%
{SmatGen1}).  Our
consideration follows closely our analysis of the simpler case of one
impurity in a Luttinger liquid wire. \cite{Aristov2009} In comparison with this
previous case, we now have a few channels of interaction in the Hamiltonian (%
\ref{eq:Ham2}). 
For the reader's convenience, we now outline the basics of our approach (see also Sec.\ \ref{sec:perturb1}). 

Consider the causal Green's function for fermions $\mathcal{G}_{ij}(x,t;y,t')= -i\langle T_{t} \psi_{j,in}(x,t)  \psi^\dagger_{j,out}(-y,t') \rangle$. In the non-interacting system ($g_{j}=0$) and given our model assumption of equal Fermi velocities in the wires, we have $  \mathcal{G} _{ij}(x,t;y,t') = \mathcal{G}  (t'-t, y-x) S^{\dagger} _{ji}$, according to (\ref{def:scatstates}). Here the scalar quantity $\mathcal{G} $ describes the kinetic part of the Green's function in the scattering states representation.   For the Wick-rotated (imaginary time) quantity $\mathcal{G}$ we have
\[
\mathcal{G}(i \omega \;;\; z)=  - i v_{F}^{-1} \mbox{sign}
(\omega)\Theta (\omega z) e^{-\omega z/v_{F}} \,.
\]

The renormalization of the junction is obtained by considering the d.c.\ limit of $\mathcal{G}_{lm}(x < - L  ,y > L,\omega\to+0 )$,  and the contribution of interaction terms order-by-order.  Without interaction we have in this limit $-i S^{\ast}_{lm}$, when we set $v_{F}=1$.   The renormalized $S$-matrix is then defined by $-iS^{\ast} _{lm} \rvert_{r}=\lim_{\omega\to+0}\mathcal{G}_{lm}(\omega;x < - L  ,y > L)$  The first correction to $\mathcal{G}_{lm}(x,0; y,t)$ in the basis of scattered states is of the form 
\begin{equation*}
\begin{aligned}  \delta\mathcal{G}_{lm}(x,t;y,t')&=2\pi g_{k}\int dz\int d\tau
 \langle\psi_{l}(x,0) \\ & \times \left( \psi^\dagger_{i'} (-z,\tau) \rho_{k}^{i'i''}  \psi_{i''}(-z,\tau)  \right)
  \\ & \times
 \left(   \psi^\dagger_{j'} (z,\tau) 
  \tilde \rho_{k}^{j'j''}  \psi_{j''}(z,\tau)   \right)
 \psi^\dagger_{j}(y,t) \rangle  S^{\ast}_{jm} \,,
\end{aligned}
\end{equation*}
where summation over repeated indices is implied, and the matrices $\rho_{k}$ and $\tilde \rho_{k}$ are given in (\ref{rhos2lambdas}). 
The two possible ways of contraction of the fermion operators lead to the expressions 
\begin{equation*}
\begin{aligned}  &2\pi g_{k}
\mathcal{G} (\tau , -z-x) 
 \rho_{k}^{lj'} \mathcal{G} (t'=0 , 2z) 
  \tilde \rho_{k}^{j'j}  
  \mathcal{G} (t-\tau , y-z)    S^{\ast}_{jm} \,, 
\end{aligned}
\end{equation*}
and 
\begin{equation*}
\begin{aligned}  &2\pi g_{k}
\mathcal{G} (\tau , z-x) 
  \tilde \rho_{k}^{li'} \mathcal{G} (t'=0 , -2z) 
 \rho_{k}^{i'j}  
  \mathcal{G} (t-\tau , y+z)    S^{\ast}_{jm} \,.
\end{aligned}
\end{equation*}
Multiplying these expressions by $e^{i \omega t}$  and integrating over $\tau, t$, we  
note that the dependence on $x$  and $y$ disappears in the limit $\omega\to0$. 
The renormalized value of $-i S^{\ast}_{lm}$ is hence of the form 
\begin{equation*}
\begin{aligned}  
-i S^{\ast}_{lm} \rvert_{r} &= -i S^{\ast}_{lm} - 2\pi g_{k}\int dz\, \Big(
 \rho_{k}^{lj'} \mathcal{G} (t'=0 , 2z)    \tilde \rho_{k}^{j'j} 
  \\ & + 
 \tilde \rho_{k}^{li'} \mathcal{G} (t'=0 , -2z)  \rho_{k}^{i'j}  \Big) S^{\ast}_{jm} \,,
\end{aligned}
\end{equation*}
or, symbolically, 
\begin{equation*}
\begin{aligned}  
 S^{\ast}_{r} .S &=1 - 2\pi i g_{k}\int dz\, \Big(
\mathcal{G} ( 0 , 2z)   \rho_{k}   \tilde \rho_{k}  
   + \mathcal{G} ( 0 , -2z) \tilde \rho_{k}   \rho_{k} \Big)  \,.
\end{aligned}
\end{equation*}



The above correction may be interpreted as the self-energy Fock diagrams $\Sigma_{F}$, since the renormalized propagator may be represented generally in terms of the self-energy as $\Sigma$ as $\mathcal{G}_{r}(-L,L)=\mathcal{G}_{0}(-L,L)+\mathcal{G}_{0}(-L,y)\Sigma(y,z)\mathcal{G}_{0}(z,L)$, where integration over $y,z$ is implied. 
Using $G(0,2z)=-i/4\pi z$, we may write this correction
(first at $g_{3}=0$)  as  
\begin{equation}
\begin{aligned} \Sigma_{F} & = - g \int _{a_{0}}^L\frac { dz }{2z} \left( [ \rho_+ ,\tilde
\rho_+ ] + [ \rho_- ,\tilde \rho_- ]\right) \\ & = -\frac g4 \Lambda
\frac{i(\lambda_4 + \lambda_6)}{\sqrt{2}} \sin\theta \left(1 + \cos\theta
\right) \end{aligned}
\end{equation}%
with $\Lambda =\ln (L/a_{0})$. We see that the matrix Green's function receives
an off-diagonal static correction, signalling the necessity to redefine the
rotation angle $\theta $ in the representation (\ref{SmatGen1}) of the $S$-matrix. It is remarkable that the above correction may be interpreted as a change of the angle $\theta$,  $\delta
\theta $ , as it is directed along the initial vector, i.e., along $\lambda
_{4}+\lambda _{6}$. This means that in the case of simple tunneling we are allowed to consider only a single
renormalization equation for $\theta$, as opposed to a set of three RG equations for $\theta$, $\psi$  and $\gamma$.

In the presence of the interaction in the third wire, contributing a term $g_{3}\rho _{3}\tilde{%
\rho}_{3}$ in the Hamiltonian, the changes in the above expression for $\Sigma $ in first
order of the interaction are minimal. In addition to the above combination $%
g\left( [\rho _{+},\tilde{\rho}_{+}]+[\rho _{-},\tilde{\rho}_{-}]\right) $
we should add $g_{3}[\rho _{3},\tilde{\rho}_{3}]$, but since $[\rho _{+},%
\tilde{\rho}_{+}]\sim \lbrack \rho _{3},\tilde{\rho}_{3}]\sim \lbrack
\lambda _{8},\tilde{\lambda}_{8}]$ we obtain
\begin{equation}
\begin{aligned} \Sigma & = i (\lambda_4 + \lambda_6) \delta \theta /\sqrt{2}
, \\ \delta \theta & = -\tfrac 14 \Lambda \left( g \sin\theta+(g+2g_3)
\sin\theta \cos\theta \right) \end{aligned}  \label{RGeq-prec}
\end{equation}
This equation (considering it as a precursor to the RG equation) is
equivalent to Eq.(8) in \cite{Das2004} (see also \cite{Lal2002}).

In our previous work, \cite{Aristov2009} we showed that the renormalization
of the impurity in the Luttinger liquid can be analyzed within the fermionic
formalism. The change of $\theta $ induced by the interaction and calculated
in first order perturbation theory in (\ref{RGeq-prec}) may now be used to
obtain the renormalization group equation for $\theta $ in lowest order 
\begin{equation}
\begin{aligned} \frac{d \theta}  {d \Lambda } & = -\frac 14 \left( g
\sin\theta+(g+2g_3) \sin\theta \cos\theta \right) \end{aligned}
\label{RGeq-1order}
\end{equation}%
In our earlier work we showed that higher order terms in the interaction may
be summed in a systematic way, to access the strong coupling domain. In
particular, we showed that the one-loop contributions to the RG equations
for the S-matrix form a ladder series, which can be resummed by solving an
integral equation of the Wiener-Hopf type. The result of this summation
reproduces the known results obtained with the bosonization method for the
weak and strong impurity case. It is found to be universal in the sense that
it does not depend on the choice of regularization of the logarithmic
divergences in the theory. We also showed that two-loop RG contributions are
absent and the three-loop RG corrections are not universal but are not very
sizeable in the whole range for a realistic choice of model parameters.

Let us now discuss how our method can be extended to the situation of the
Y-junction. First we discuss the ladder summation, then we present the
results of a computer symbolic calculation of the three-loop terms.

\subsubsection{Ladder summation}

\label{sec:LadderSum}

Our previous solution of the ladder equation  \cite{Aristov2009} 
amounted to a the dressing of
the interaction in the presence of the impurity. It thus led to a
replacement $g\rightarrow \widetilde{g}$ in an equation analogous to (\ref%
{RGeq-1order}) with 
\begin{equation}
\begin{aligned} \widetilde g & = \frac{2 g}{1+ \sqrt{1-g^2 } - g Y}
\end{aligned}  \label{dressing1}
\end{equation}%
where $Y=Tr(\rho _{-}\tilde{\rho}_{-})$ ; notice the different sign in front
of $gY$ here, which is a result of the different parametrization of the
S-matrix in our previous work.

In view of the symmetry $1\leftrightarrow 2$, we have $Tr(\lambda_3 .
\tilde\lambda_8) =Tr(\tilde \lambda_3 . \lambda_8) =0 $. This means that the
symmetric density combinations remain orthogonal after the scattering $%
Tr(\rho_- \rho_+) = Tr(\tilde \rho_- \tilde \rho_+) = Tr(\rho_- \tilde
\rho_+) =Tr(\tilde \rho_- \rho_+) =0$. This property allows us to perform
the ladder summation separately in each of the channels, as described below.

First we consider the simpler case of $g_3=0$, the absence of interaction in
the third wire. From Eq.\ (\ref{rotL38}) we have $Tr(\rho_- \tilde \rho_-)=
\cos \theta$ and $Tr(\rho_+ \tilde \rho_+)= \tfrac14 (3+\cos2\theta) $. The
ladder summation of the one-loop RG contributions in each channel is
performed along the previous guidelines \cite{Aristov2009} and the resulting
RG equation is written as

\begin{equation}
\begin{aligned} \frac {d \theta}  {d \Lambda } \equiv \beta(\theta) &= -\frac g 4
\sin \theta \left( \frac{2}{1+d - g \cos\theta} \right. \\& + \left.
\frac{2\cos\theta}{1+d - (g/4) (3+\cos2\theta)} \right) \end{aligned}
\end{equation}%
with $d=\sqrt{1-g^{2}}$.

Let us now discuss the simultaneous presence of $g,g_{3}\neq 0$. The result
of the summation in the $\rho _{-}$ channel is the same, whereas the
channels $\rho _{+}$ and $\rho _{3}$ begin to mix in higher orders of the
interaction. We omit the details of the derivation below and only provide
intuitive arguments for the result obtained.

The dressing of the interaction occurs in the individual wires,
characterized by two density components $\rho_+ , \rho_3$, which are
orthogonal before the scattering, $Tr(\rho_{+} \rho_{3})=0$. If these
density components were orthogonal to those \emph{after the scattering}, ($%
Tr(\rho_{+} \tilde \rho_{3}) = Tr(\rho_{+} \tilde \rho_{+}) = 0$, etc.) then
the dressed interaction would simply be given by (\ref{dressing1}) with $Y=0$%
. We define renormalized interaction constants in the form of two "charges" $%
q, q_3$ by $g \to 2 g/(1+ \sqrt{1-g^2 }) = 2 q^{-1}$, $g_{3} \to 2 g_{3}/(1+ 
\sqrt{1-g_{3}^2 }) = 2 q_{3}^{-1}$ and introduce the diagonal matrix $\hat Q
= diag[q, q_3] $ here.

According to Eq.\ (\ref{rhos2lambdas}) the two density components $\rho _{+},\rho _{3}$ are
connected to a vector of scattering eigenmodes, $\phi =(\lambda _{0},\lambda
_{8})/\sqrt{2}$, by 
\begin{equation}
\begin{aligned} \begin{pmatrix} \rho_+ \\ \rho_3 \end{pmatrix} &=
\frac1{\sqrt{2}} \hat U \begin{pmatrix} \lambda_0 \\ \lambda_8 \end{pmatrix} \,,
\\ \hat U &= \frac1{\sqrt{3}} \begin{pmatrix} \sqrt{2}, & 1 \\ 1, &
-\sqrt{2} \end{pmatrix} \,,  \end{aligned}   \label{rho3+lambda08}
\end{equation}%
with $\hat{U}^{2}=1$ and $Tr(\phi _{i}\phi _{j})=\delta _{ij}$. Notice that
the modes $\lambda _{0}$ and $\lambda _{8}$ remain orthogonal both before
and after the scattering, which is reflected in the matrix $\hat{Y}=Tr(\phi
_{i}\tilde{\phi}_{j})=diag[1,\tfrac{1}{4}(1+3\cos 2\theta )]$. The result of
the ladder summation is represented by the matrix $(\hat{Q}-\hat{U}.\hat{Y}.%
\hat{U})^{-1}$.

The renormalization of the $S$-matrix in the first order occurs due to the
non-vanishing commutators of densities $[\rho_{+}, \tilde \rho_{+}]$, $%
[\rho_{3}, \tilde \rho_{3}]$. Going to higher orders in the interaction, we
have to take into account also the mixed commutators, $[\rho_{+}, \tilde
\rho_{3}]$, $[\rho_{3}, \tilde \rho_{+}]$. Noticing that the only
non-commuting components here are $\lambda_8$ and $\tilde \lambda_8$, and
using Eq.\ (\ref{rho3+lambda08}), we find that the effect of renormalizing
the interaction amounts to the replacement

\begin{equation}
(g+2g_{3})\sin 2\theta \rightarrow 6 \left[ \hat{U}.\hat{Q}.\hat{%
U}-\hat{Y}\right] _{22}^{-1} \sin 2\theta  \;,
\end{equation}%
which leads to the explicit result for the $\beta $-function in the ladder
approximation: 
\begin{equation}
\begin{aligned} \beta_L(\theta) & = -\frac 1 2 \left( \frac{ 2\sin
2\theta}{Q -\cos2\theta} + \frac{\sin \theta }{ q - \cos\theta} \right) \,,\\
Q& = \frac{4qq_3 - 2 q -3 q_3+ 1}{ 2q + q _3 -3 } \,,\\ q& = \frac{1+K}{1-K}
,\quad q_3 = \frac{1+K_3}{1-K_3} \,,\\ K &=\sqrt{\frac{1-g}{1+g}}, \quad
K_{3}=\sqrt{\frac{1-g_{3}}{1+g_{3}}} \,, \end{aligned}  \label{RGeq-38}
\end{equation}%
where we also provide the alternative definition of $q,q_{3}$ through the
Luttinger parameters $K,K_{3}$.

The equation (\ref{RGeq-38}) is a central result of this paper,
and we analyze it  in some detail in the next subsection. It should be noted 
that in the limit $g_{3} \to 0$ and for small $g$ this equation reduces to the Eq.\ (1)
of Ref.\   [\onlinecite{Aristov2010}], where the fixed points of the point contact model
were discussed.  We confirm Eq.\ (\ref{RGeq-38})  
by calculating the perturbative corrections to the $S$-matrix up to the
third order of the interaction in Sec.\ \ref{sec:perturb1}.

\subsubsection{\label{sec:ExplicitLadder}Explicit solution of the RG equation}

Let us analyze the RG equation (\ref{RGeq-38}). Introducing the variable $%
x=\cos \theta $, which determines the conductivity components via (\ref%
{Gmatrix2}), we rewrite it in the form

\begin{equation}
\begin{aligned} \frac{dx}{d\Lambda } &=\frac{1}{2}(1-x^{2})
\left(\frac{4x}{Q+1-2x^{2}}+\frac{1}{q-x} \right) \,,\\ & =
\frac{3}{2}(x^{2}-1)\frac{(x-x_{3})(x-x_{4})}{(x^{2}-q_{0})(x-q)}\,,
\end{aligned}  \label{explRGladder}
\end{equation}%
with 
\begin{equation}
\begin{aligned}
q_{0}&=\frac{1}{2}(Q+1)\,, 
\\  x_{3,4}&=\frac{1}{3}[q\mp \sqrt{3q_{0}+q^{2}}]\,,
\end{aligned}  \label{fixedpoints}
\end{equation}%

Here $Q$ is defined in (\ref{RGeq-38}). The zeros of
the r.h.s. , the four fixed points (FPs) of the RG equation, are given by $%
x_{1,2}=\pm 1$, and $x_{3,4}$ . For later reference we label the fixed points $x_{1},x_{2},x_{3},x_{4}$ as $N,A,M,Q$. Fixed point $N$ corresponds to $G_{a}=G_{b}=0$, i.e the three wires are separated. At fixed point $A$ we have $G_{a}=1,G_{b}=0$, i.e. the third wire is disconnected whereas the main wire is perfectly conducting. The fixed points $M,Q$ are located on the boundary of the physically accessible region, unless they are outside the region defined by (\ref{bound}).

For repulsive interaction, $0<K,K_{3}<1$, the fixed point $x_{4}>1$ is
outside the physical domain, $|x|\leq 1$. One can easily verify that the two
fixed points $N,A$ at $x_{1,2}=\pm 1$ are stable, i.e.\ $d\beta /dx|_{x_{1,2}}<0$. We will see below that $A$ turns unstable in the general symmetric case.
Fixed point $M$ at $x_{3}\in (-1,0)$ is unstable. At weak coupling,
we write $K=1-g $ and $K_{3}=1-g _{3}$ with $0<g^{2} \ll g
,g _{3}\ll 1$, and get $x_{3}\simeq -{g }/{(g+2g _{3})}$, cf.\ \cite{Aristov2010}.  

The equation (\ref{explRGladder}) may be integrated to give the implicit solution for 
$x$ as a function of the length $L$\ in the form 
\begin{equation}
\begin{aligned}
F[x(L)] & =(1-x)^{\gamma_{1}}(1+x)^{\gamma _{2}}|x-x_{3}|^{\gamma
_{3}}|x-x_{4}|^{\gamma _{4}}
\\ & =F[x(L_{0})]\, L/L_{0} , 
\end{aligned}
\label{explicit}
\end{equation}
where 
\begin{equation}
\begin{aligned}
\gamma _{1} &= \frac{1}{3}\frac{(1-q_{0})(1- q)}{(1 - x_{3})(1 - x_{4})} \\ & =
- \left( K^{-1}+K_{3}^{-1} -2 \right)^{-1} , \\
\gamma
_{2} & =\frac{1}{3}\frac{(1-q_{0})(1 + q)}{(1 + x_{3})(1 + x_{4})} \\ &
= - \left( \tfrac12 ( K^{-1} +K - 2)+K_{3}^{-1} -1 \right)^{-1} , \\
\gamma _{3} & =\frac 23
\frac{q-x_{3}}{x_{4}-x_{3}}\frac{q_{0}-x_{3}^{2}}{1-x_{3}^{2}}, \\
\gamma _{4} & =- \frac 23
\frac{q-x_{4}}{x_{4}-x_{3}}\frac{q_{0}-x_{4}^{2}}{1-x_{4}^{2}} ,  
\end{aligned}  \label{indices}
\end{equation}

We see that the fixed point $N$  at $x=1$, corresponding to three fully detached
wires, is characterized by the exponent $\gamma _{1}$, which is determined by
the sum of two boundary exponents, $K^{-1}$ and $K_{3}^{-1}$. The second
fixed point $A$ at $x=-1$, corresponding to the ideal wire 1 and 2 and the
detached wire 3, relates to the exponent $\gamma _{2}$, which is governed by the
boundary exponent of the third wire, $K_{3}^{-1}$, and the bulk anomalous
dimension of the fermion operator $(K^{-1}+K)/2$, the latter quantity
defining the local density of states in tunneling experiments.

In the important special case of arbitrary but equal interaction strength, $K_{3}=K$ we
have $Q=(4q-1)/3$, $q_{0}=(2q+1)/3=x_{4} $ and $x_{3} = -1/3$. The
expressions for $\gamma _{3,4}$ simplify and we have 
\begin{equation}
\begin{aligned} \gamma _{3}^{-1} & = 6\frac{1-K}{(2+K)^{2}} , \\
& = \tfrac 23 (K^{-1}-1) - \tfrac23 \tfrac{4-K}{(2+K)^{2}} 
\left( K^{-1}+K -2 \right),
\\ \gamma
_{4}^{-1} & = -6\frac{3-K}{K(3+K)} , \end{aligned}
\label{exponents-sym}
\end{equation}

For completeness, we also consider the case of attractive interaction $K,K_{3}>1$. 
First of all, the role of the fixed points is reversed, so that
the fixed points $N,A$ at $x=\pm 1$ are unstable and the third fixed point $M$ at $x_{3}$
becomes stable. Further, we observe that while for repulsive interaction $%
x_{4}>1$ always lies outside the physical region of $x$, this is different
in the attractive case. For simplicity, let us first consider the case of equal
interaction in the wires, $K=K_{3}$.  We have $x_{3}=-1/3$ 
and the value of $x_{4}=(2q+1)/3=\tfrac{3+K}{3(1-K)}$  reaches the physical 
range of $x$ first and coincides with $x=-1$
at $K=3$. At this point the inverse exponents $\gamma _{2}^{-1}=\gamma
_{4}^{-1}=0$. A subsequent increase of $K$ takes $-1<x_{4}<-1/3$ inside the
physical domain. One can verify that the fixed point $Q$ at $x_{4}$ in this latter case is
unstable, whereas the fixed  point $A$ at the edge $x=-1$ becomes stable. 

Another interesting possibility for attractive interaction is merging of the points $x_{3}$ 
and $x_{4}$. It happens at $3q_{0}=-q^{2}$ in (\ref{fixedpoints}), or at 
\[ K_{3}=2K\frac{1+K^{3}}{(1+K)^{2}(2K-1)} \,,\]
and the position of FP in this case is 
\[ x_{3} = x_{4} = \tfrac13 {(1+K)}/{(1-K)} \,.\] 
The last equations show, that in order to have $|x_{3,4} |<1$ one should let $K\ge 2$,
$K_{3}\ge4/3$. We return to these
questions below in Sec.\ \ref{sec:discussion}, when discussing the ``phase diagram'' 
of our model.

\subsection{\label{sec:GenSym} General Y junction}

Let us now analyze the general case of the $S$-matrix, Eq.\ \ref%
{SmatGen1}. As shown above in Sec.\ \ref{sec:RGtun1st}, the RG flow to first-order in
the interaction is determined by two contributions, which are proportional
to $[\tilde{\lambda}_{3},\lambda _{3}]$ and $[\tilde{\lambda}_{8},\lambda
_{8}]$. We write 
\begin{equation}
S^{-1}.S_{r}-1=\Lambda (\alpha _{3}[\tilde{\lambda}_{3},\lambda _{3}]+\alpha
_{8}[\tilde{\lambda}_{8},\lambda _{8}])\,,
\label{RG:dS}
\end{equation}%
where $\alpha _{3}=-g/4,\alpha _{8}=-(g+2g_{3})/12$ are introduced for
brevity. Comparing the right-hand side of the above expression with the
parametrization (\ref{genSparam1}) of the S-matrix, we can find the
corresponding change of the parameters $\tilde{\theta},\phi ,\tilde{\psi}$
in the renormalized scattering matrix $S_{r}$.

After some calculation we find the following set of RG equations  
\begin{equation}
\begin{aligned} \frac{d \tilde\theta}{d\Lambda} &= \alpha_{3} (\cos
\tilde\psi \sin \tilde\theta + \sin \tilde\psi \cos \tilde\theta \sin\phi)
\\ & + 3 \alpha_{8} \cos \tilde\theta \sin \tilde\theta \cos^{2} \phi , \\
\frac{d\phi}{d\Lambda} & = \alpha_{3} \frac{\cos \phi \sin \tilde\psi}{\sin
\tilde \theta} - 3 \alpha_{8} \sin\phi \cos \phi , \\ \frac{d
\tilde\psi}{d\Lambda} & = 3 \alpha_{3} (\cos \tilde\psi \sin \tilde\theta
\sin\phi + \sin \tilde\psi \cos \tilde\theta )\,. \end{aligned}  \label{difeq1}
\end{equation}%
At first sight, there are three equations for three parameters. However, one
can check that ${d}\widetilde{\psi }/{d\Lambda }=3\cos ^{2}\widetilde{\theta 
}{d(\sin \phi \tan }\widetilde{\theta }{)}/{d\Lambda }$, irrespective of the
interaction, so that only two of the equations are independent.

Indeed, using the equivalent parametrization of the $S$-matrix in the form
Eq.\ (\ref{genSparam2}), we obtain the RG equations in the new variables $%
\theta ,\psi ,\gamma $ as : 
\begin{equation}
\begin{aligned} \frac{d\theta}{d\Lambda} &= \sin \theta ( \alpha_{3} \cos
\psi + 3\alpha_{8} \cos \theta ) , \\ \frac{d \psi}{d\Lambda} & = \alpha_{3}
\sin \psi \left(3 \cos\theta + \frac1 {\cos \theta} \right) , \\
\frac{d\gamma}{d\Lambda} & = \alpha_{3} \frac {\sin \psi}{ \cos \theta }
\end{aligned}  \label{difeq2}
\end{equation}%
Therefore, we have only two independent RG equations (\ref{difeq2}) for the
variables $\theta $ and $\psi $, which define the components of the
conductance matrix (\ref{Gmatrixhat}). The third component $\gamma $ is not
independent and is determined by $\theta $ and $\psi $ ; it does not enter
the conductance.

It is also instructive to express the RG equations directly in terms of the
components of the  conductance tensor (\ref{defGprime}). We introduce 
\begin{equation}
\begin{aligned} a & = \tfrac12 Tr(\tilde\lambda_{3} \lambda_{3}) = \cos
\theta \cos \psi, \\ b & = \tfrac12 Tr(\tilde\lambda_{8} \lambda_{8}) =
\tfrac14 (1+3\cos 2\theta) , \end{aligned}  \label{def:ab}
\end{equation}
so that the conductance matrix is $\hat{G}=diag(\tfrac12 (1-a),\tfrac23 (1-b))$. From (\ref{difeq2}%
) we find 
\begin{equation}
\begin{aligned} \frac{d a }{d\Lambda} & = - 2 \alpha_{3} (1+b-2a^{2}) - 2
\alpha_{8} (1-b)a , \\ \frac{d b}{d\Lambda} &= - 2 (1-b) ( \alpha_{3} a +
\alpha_{8} (1+2b) ) , \end{aligned}  \label{difeq3}
\end{equation}
or, directly in terms of conductance, 
\begin{eqnarray}
\frac{d G_{a} }{d\Lambda} & = & \alpha_{3} \left(
8G_{a}(1- G_{a}) - \tfrac32 G_{b} \right) + 
\tfrac 32 \alpha_{8} G_{b}(1-2G_{a}) , \nonumber \\ 
\frac{d G_{b}}{d\Lambda} &=& 2  \alpha_{3} G_{b}(1-2G_{a}) +
6 \alpha_{8} G_{b}(1-G_{b}) ,   \label{difeq4}
\end{eqnarray}

Comparing Eq.\ (\ref{difeq2}) to the above special case, ($\psi = \gamma =0$), 
we observe that the arguments, leading to the possibility of the ladder
summation in Sec.\ \ref{sec:LadderSum}, remain valid. Hence we can use the
previous result, which amounts to the substitution 
\begin{equation}
\begin{aligned} \alpha_{3} & = - \frac12 \frac {1} {q - a}
= - \frac12 \frac {1} {q -1 + 2G_{a} } , \\ \alpha_{8}
&= -\frac12 \frac 1 {Q_{1} - b} = -\frac12 \frac 1 {Q_{1} -1+ \tfrac32 G_{b}}
, \\ Q_{1}& = \frac{3qq_3 - q -2 q_3}{ 2q +
q _3 -3 } \end{aligned}  \label{def:alphas}
\end{equation}
with $q,q_{3}$ defined in (\ref{RGeq-38}). The equations (\ref{difeq3}), (\ref{difeq4}), 
(\ref{def:alphas}) are the main result of this paper.

Notice that the RG equation for the quantity 
\begin{equation}
\begin{aligned} G_{\perp}  & = G_{b} +(2G_{a}-1)^{2}  \,, \\ 
&= \tfrac23 (1-b) +a^{2}  =  1-\cos^{2}\theta \sin^{2}\psi \,,
\end{aligned}  \label{def:Gperp}
\end{equation}
due to Eq.\ (\ref{difeq4}) has a form
\begin{equation}
\frac{d G_{\perp} }{d\Lambda} \propto (1-G_{\perp})
\label{GperpRG}  
\end{equation}
which, together with (\ref{bound}) and second line in (\ref{difeq4}), shows that the boundaries for the observable conductances for free fermions,  $G_{\perp}=1$  and $G_{b}=0$, are the RG fixed lines in the interacting case. 

Concerning the character of the fixed points in this more general
situation, there is a qualitative change. In the tunneling case, and for repulsive
interaction,  we had
two stable FPs , $A,N$, corresponding to i) one wire detached ($x=-1$) and ii) three
wires detached ($x=1$). The unstable FP $M$ ($x_{3}$) was located between these
two limiting cases.

Now we have two independent components of the
conductance. We find only one truly stable FP : $N$, the case
of all three wires detached, $G_{a}=G_{b}=0$. The previous stable FP $A$ with
one wire detached ($G_{a}=1,G_{b}=0$) transmutes into a saddle point, so it
is not truly stable. The third (unstable) FP $M$ is at the boundary of the 
region of conductances in the $G_{a}-G_{b}$-plane allowed by unitarity and
is unstable in both directions.

\begin{figure}[tbp] 
\includegraphics*[width=0.9\columnwidth]{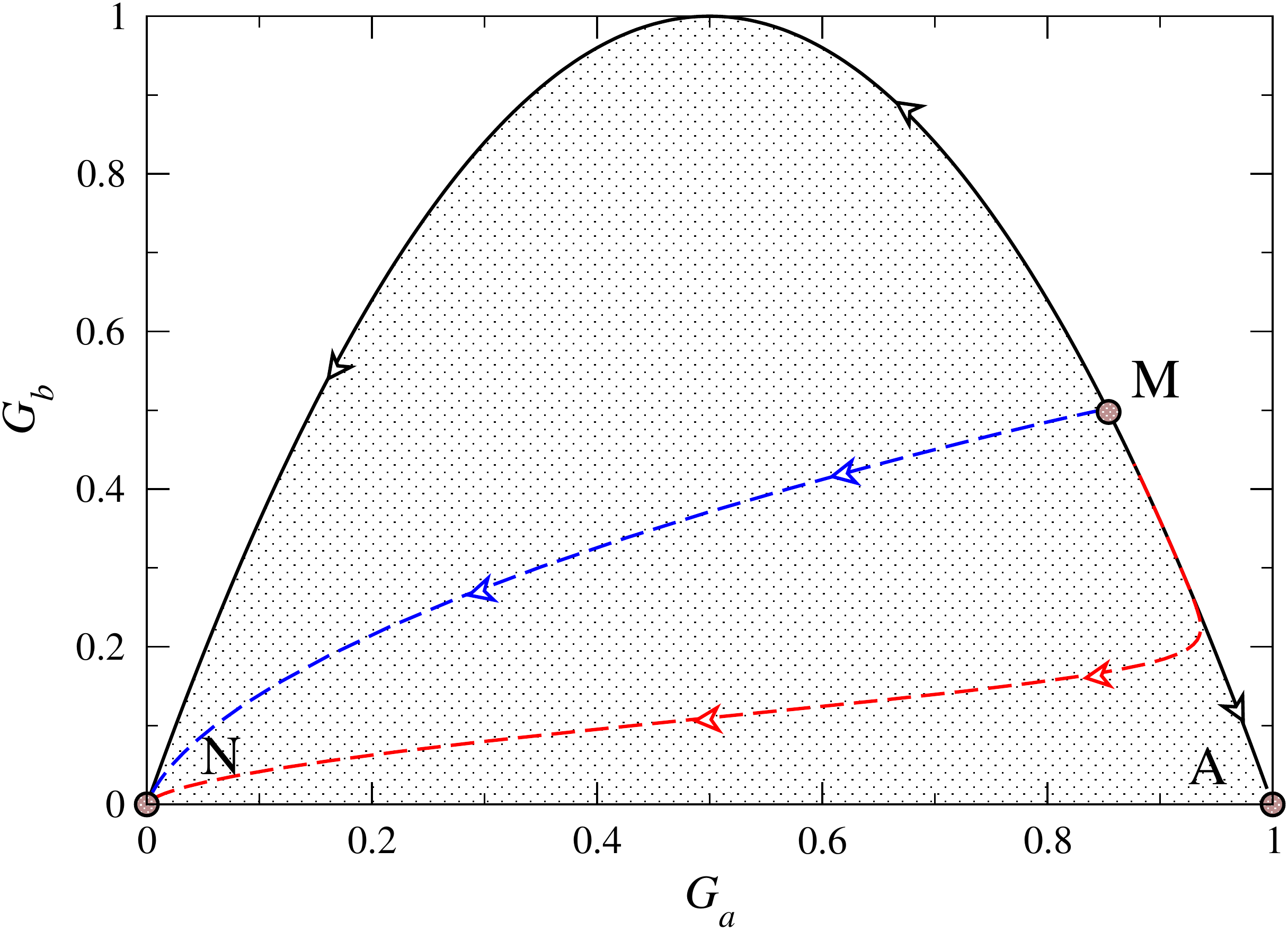}
\caption{(Color online) Ladder approximation for the $\protect\beta$%
-function and RG flows for $g=0.2$ and $g_{3}=0.03$. Three fixed points are
shown by dots and are : (i) stable one at $G_{a}=G_{b} =0$, (ii) unstable
one at $G_{a}\simeq 0.8542$, $G_{b}\simeq 0.4979$, and (iii) saddle-point
type fixed point at $G_{a}=1$, $G_{b}=0$. The values of conductances,
allowed by unitarity, are shown as a shaded region. }
\label{fig:RGflow}
\end{figure}

The RG flows for repulsive interactions of strength $g=0.2,g_{3}=0.03$ are depicted 
in Fig.\ \ref{fig:RGflow}. It is interesting to note that the three representative flow 
trajectories emanating from fixed point $M$ indicate nonmonotonic behavior of $G_{b}$ 
(black) or of $G_{a}$ (red) as a function of, e.g. temperature. As shown in 
\cite{Aristov2010} this behavior at $g_{3}=0$ appears first in second order perturbation 
theory in the interaction $g$.  
It was demonstrated there, that the intermediate $M$ point at $g_{3}=0$ appears due to 
the competition between the bulk zero-bias anomaly and the scattering off Friedel 
oscillations induced by the third wire. This underlines the difference in the origin of the 
intermediate point in Ref.\   [\onlinecite{Aristov2010}] and the origin 
of the intermediate point in this and other studies \cite{Lal2002, Chen2002, Das2004, Barnabe2005, Oshikawa2006} with $g_3\neq 0$.

\begin{figure}[tbp] 
\includegraphics*[width=0.9\columnwidth]{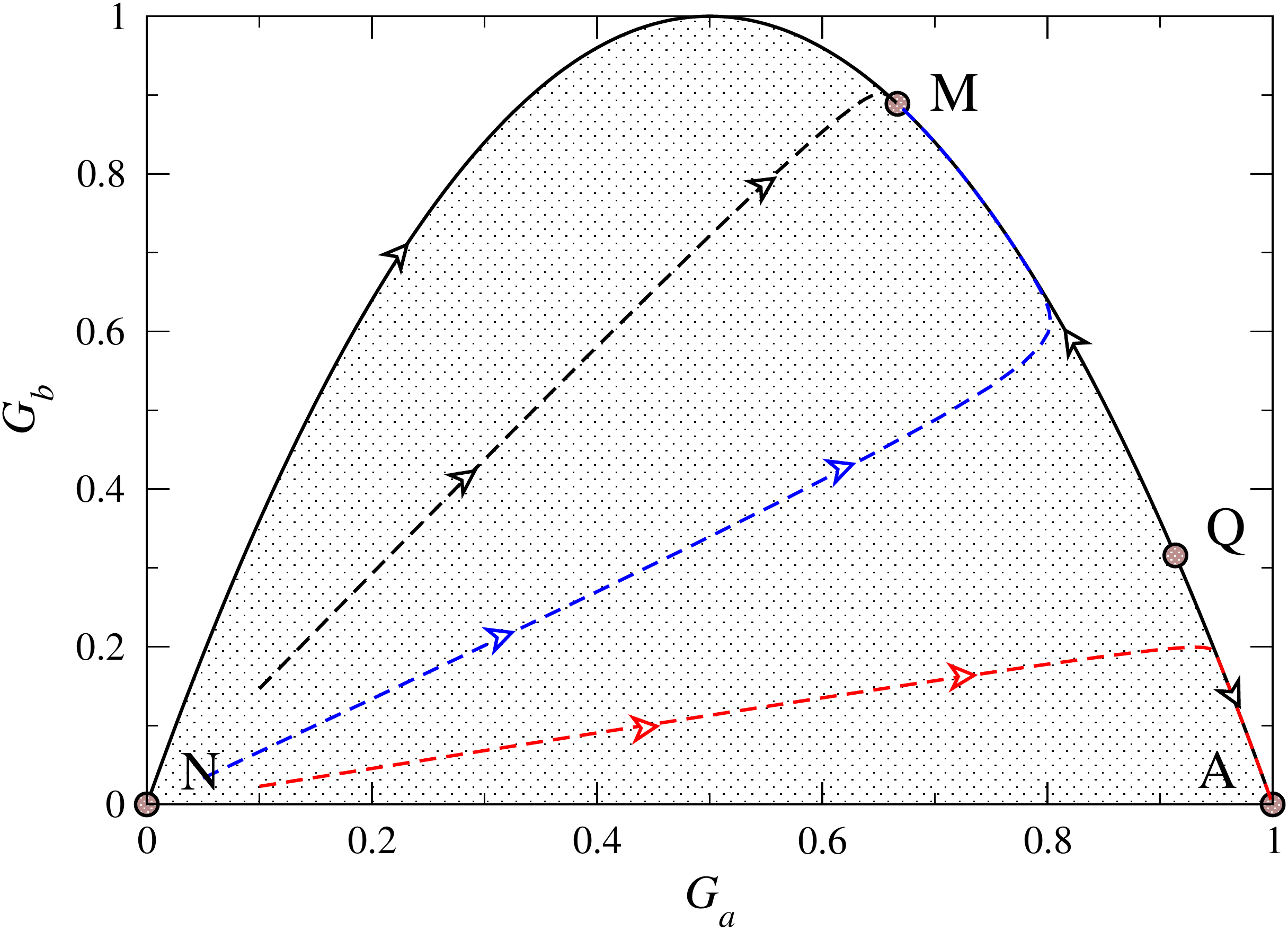}
\caption{(Color online) Ladder approximation for the $\protect\beta$%
-function and RG flows for $K=K_{3} = 3.7$. Four fixed points are
shown by dots and are : (i) unstable one at $G_{a}=G_{b} =0$, (ii) stable
one at $G_{a}=1$, $G_{b}=0$, (iii) stable one at $G_{a}=2/3$, $G_{b}=8/9$ and (iv) 
new (appearing at $K>3$) unstable fixed point at $G_{a}\simeq0.9136$, $G_{b}\simeq0.3159$. 
The values of conductances,
allowed by unitarity, are shown as a shaded region. }
\label{fig:RGflowK37}
\end{figure}

The scaling behavior in the case of attractive interaction is even more interesting (although attractive effective interaction is not easily realized in nature!). The conductances at $M$ are given by $G_{a}=(1-x_{3})/2$ and  $G_{b}=1-x_{3}^{2}$, where $x_{3}$ is given in (\ref {fixedpoints}). As mentioned above, fixed point $N$ becomes unstable, while $M$ is stable. In the case of not too strong attractive interaction, $1< K, K_{3} <3$ , $M$ is the only stable fixed point. Again the behavior of both, $G_{a}$ and $G_{b}$ may be nonmonotonic. For even stronger attractive interaction the fourth fixed point $Q$ enters the physically accessible region as a further unstable FP. This leads to a switch of fixed point $A$ from unstable to stable. In Fig.\ \ref{fig:RGflowK37} the flow diagram is shown for strong attractive interaction, $K=K_{3}=3.7$ . 

It is useful to analyze the RG equations (\ref{difeq3}) around the $N$ and $A$ fixed points. 
For the $N$ point which corresponds to $a=b=1$ we write $a=1-a'$, $b=1-b'$ and find in the leading order of the small deviations $a'>0$, $b'>0$: 
\begin{equation}
\begin{aligned} 
\frac{d }{d\Lambda} \left(a'-\tfrac13 b' \right)& \simeq 2  \left(1-K^{-1} \right)
\left(a'-\tfrac13 b' \right) \, , \\ 
\frac{d }{d\Lambda} b'& \simeq   \left(2-K^{-1}-K_{3}^{-1} \right) b'
\,. \end{aligned}  
\label{RGeq:N}
\end{equation}
Similarly, around the $A$ with $a=-b=-1$ we expand $a=-1+a'$, $b=1-b'$ to obtain:
\begin{equation}
\begin{aligned} 
\frac{d }{d\Lambda} \left(a'-\tfrac13 b' \right)& \simeq 2  \left(1-K \right)
\left(a'-\tfrac13 b' \right) \, , \\ 
\frac{d }{d\Lambda} b'& \simeq   \left(1-K_{3}^{-1}- \frac{(1-K)^{2}}{2K} \right) b'
\,. \end{aligned}  
\label{RGeq:A}
\end{equation}
In the equations (\ref{RGeq:N}), (\ref{RGeq:A}) the combination $a'-\tfrac13 b'>0$ measures the distance to the parabola $b=(3a^{2}-1)/2$ of the simple tunneling case $\psi=0$ in (\ref{def:ab}). Thus we confirm that the border lines of the physical sector of conductances, 
Eq.\  (\ref{bound}), are the ``fixed lines'' of our RG equations. 

The meaning of scaling exponents in (\ref{RGeq:N}), (\ref{RGeq:A}) is as follows. The quantity $a'-\tfrac13 b'$ is the difference of the Y-junction from the simple tunneling case which may arise primarily due to impurity scattering located in the main wire. It is then quite natural that it scales at $N$ point with the usual double exponent, $2(1-K^{-1})$, of the weak tunneling between two equal Luttinger liquids. The quantity $b'$ stands for the tunneling conductance and its exponent combines the weak tunneling contributions from the main and the third wire.  At the fixed point $A$ we have the conductance $G_{b}=\tfrac23 b'$ defined by the weak tunneling from the third wire, $(1-K^{-1}_{3})$, into a perfect Luttinger liquid with a bulk exponent $(1-K)^{2}/{2K}$.  The small value of $a'-\tfrac13 b'$ in this case corresponds to a weak barrier in the main wire, which scales with the exponent $2(1-K)$. \cite{Kane1992, Aristov2010}

\subsection{\label{sec:FullySym}Fully symmetric Y junction}

In this section we consider the case of full symmetry between the wires, which 
means equal strength of interactions, $K=K_{3}$. The asymmetry at the Y-junction may 
still lead to an asymmetry in the physical properties of our system. However, it is a  remarkable fact that the symmetric Y-junction, $a=b$, in other terms $G_{b}=\tfrac43G_{a}$ , remains a fixed line in this case. 

We observe that for $K=K_{3}$ we have $q=q_{3}=Q_{1}$ and $ \alpha_{3} = \tfrac12
(a-q )^{-1}$, $ \alpha_{8} = \tfrac12 (b-q )^{-1}$ in (\ref{def:alphas}). Introducing the symmetrized variables, $c=b-a$, $\bar a=(a+b)/2$, we obtain from  Eq.\ (\ref{difeq3}) 
\begin{equation}
\begin{aligned} 
\frac{d \bar a }{d\Lambda} & =  \frac {(1-\bar a) (1+3\bar a)}{q-\bar a} + {\cal O}(c^{2})   \,, \\ 
\frac{d c}{d\Lambda} &=  \frac{1+3\bar a^{2}-4\bar aq}{(q-\bar a)^{2}} c  
+ {\cal O}(c^{2})
\,. \end{aligned}  \label{difeqFullsym}
\end{equation}
This set of equations shows that the RG flows form a ``spindle'' shape around the line
$a=b$,  with two edge FPs at  $a=b=1$ (N point) and $a=b=-1/3$ (M point). 
The first equation  in (\ref{difeqFullsym}) is easily integrated with the result (cf.\ 
(\ref{explicit}))
\begin{equation}
\tilde F[\bar a(L)]= (1-\bar a)^{\gamma_{2}}\left(\tfrac13+\bar a\right)^{\gamma'_{3}}
=\tilde F[\bar a(L_{0})] \,L/L_{0}
\end{equation}
here $\gamma_{2}^{-1}=-2(K^{-1}-1)$ 
follows from (\ref{indices}) and 
\begin{equation} 
\left(\gamma'_{3} \right)^{-1}  = 6\frac{1-K}{(2+K)} \,.
\label{expMpoint}
\end{equation} 
Notice that the similar-looking $\gamma_{3}$ in (\ref{exponents-sym}) defines the scaling exponent for $c$ in (\ref{difeqFullsym}).

\section{\label{sec:perturb1}Perturbation theory and non-universal terms}

The ladder summation discussed above captures contributions of a certain type (one-loop) to all orders in the interaction. By its structure it is similar to the RPA summation scheme, which works perfectly well for the description of the bulk properties of a Luttinger liquid, due to the absence of multi-tail fermionic loops in the linearized dispersion model \cite{Aristov2007}. There remains the question of the importance of higher loop contributions. In our work on the Luttinger liquid with barrier we were able to show in perturbation theory that the two-loop contribution vanishes and the three-loop contribution is subdominant in the neighborhood of the fixed points. As a result, the exponents of the power laws of the conductance in length $L$ or temperature $T$ turned out to be in exact agreement with those obtained by methods of bosonization, and the thermodynamic Bethe ansatz. In this section we explore to which extent a similar result is true in the case of a Y-junction.

\subsection{Simple tunneling case}

The symbolic computation of the diagrams up to the third order can be
performed as explained in the previous work. \cite{Aristov2009} The
diagrammatic rules are similar, with the only difference that we have $%
3\times3$ matrices for the vertices now. This means that the kinematic
structure for each diagram is the same both for the $Y$-junction and for the
barrier in the Luttinger liquid. Therefore we can use the previously
obtained results for the individual diagrams, while supplying them with
different matrix prefactors.

We may represent the renormalized scattering matrix, $S_{r}$, in the form 
\begin{equation}
S^{\dagger }.S_{r}=1+g\Sigma _{1}+g^{2}\Sigma _{2}+g^{3}\Sigma _{3}+\ldots ,
\label{Smatexpa}
\end{equation}%
(in the presence of both $g$ and $g_{3}$ we consider $\Sigma _{j}$ as
dependent on the ratio $g_{3}/g$). In the next step, we verify the
functional form of the matrix $S^{\dagger }.S_{r}=\exp i\delta \theta
(\lambda _{4}+\lambda _{6})/\sqrt{2}$ and determine the scalar quantity $%
\delta \theta $ as a function of $g,\theta ,\Lambda ,g_{3}/g$. We have for
the renormalized value $\theta _{r}=\theta +\delta \theta (\theta )=f(\theta
,g\Lambda ,g)$ and inverting this equality we may write $\theta =f^{\ast
}(\theta _{r},g\Lambda ,g)$. Demanding now that the initial "bare" value $%
\theta $ should not depend on $\Lambda $, we may derive the RG equation. 
\cite{Aristov2009}

Following these steps, we arrive at an expression for the $\beta$-function,
which we keep up to third order in $g, g_{3}$. We compare this perturbative
result with the Taylor expansion of Eq.\ (\ref{RGeq-38}) and confirm the
identity of the two expressions up to second order in the interaction. In
the third order of $g$, the direct calculation of the $\beta$-function
provides the ladder contribution given by the corresponding term in the
Taylor expansion of Eq.\ (\ref{RGeq-38}) and in addition a three-loop
contribution to $\beta(\theta)$ beyond the ladder result. In accordance with 
Ref.\ [\onlinecite{Aristov2009}], we find that this three-loop correction is a subleading
contribution, stemming from the non-universal parts of the diagrams. As was
explained in the previous work, performing the evaluation of the diagrams we
find a few generic integrals, which contain non-universal numbers $b_2, b_3,
b_4$ which for $T=0$ are equal to $\pi^2/6, 2\ln 2, 2(\ln2 -1)$,
respectively. \cite{misprint} The extra term in the $\beta$-function, which should be added
to (\ref{RGeq-38}) is given by

\begin{equation}
\begin{aligned} \beta_3(\theta) & = \frac {g^3}{128} \sin^3 \theta \left( b_2
f_2 + b_3 f_3 + b_4 f_4 \right) \label{RG_beta3} \\ f_2 & = 1+ 8 z+ 2z^2 + 4
z^3 + (1+2g_3/g)^2 \\ f_3 & = (1-z)^2 -4g_3/g \\ f_4 & = 2 (1+z) (1+ z ^2 )
\\ z(\cos \theta) &= (1+2g_3/g) \cos \theta \end{aligned}
\end{equation}%
Comparing with the case of the single wire with barrier we see that the $%
f_{j}$ are now smooth functions of $\theta $, rather than constants.

Let us consider the influence of $\beta _{3}(\theta )$ on the fixed point
values defined by $\beta _{L}=0$ of the reduced RG equation (\ref{RGeq-38}),
as analyzed in the previous subsection. Near the two fixed points $N,A$, at $\theta
=0,\pi $, corresponding to $x=\pm 1$, the term (\ref{RG_beta3}) is small,
proportional to the third power of $\sin \theta $. This means that neither
the positions of these two fixed points, nor their scaling exponents $\gamma
_{1,2}$ in (\ref{indices}) are affected by $\beta _{3}(\theta )$.

The position of the third fixed point, $x_{3}$, is defined in the ladder
approximation by $(g+2g_{3})x_{3}+g=0$, up to higher order corrections. The
latter condition reads $z(x_{3})=-1$ in (\ref{RG_beta3}), and we find
accordingly that the\ $f_{j}$ at $x_{3}$ are given by $%
f_{2}=-4(2+g_{3}/g)(1-g_{3}/g)$,  $f_{3}=4(1-g_{3}/g)$,  $f_{4}=0$ . It
follows that the position of the third fixed point $x_{3}$ is affected by
the $\beta _{3}$ term at $g_{3}\neq g$; and the critical exponent $\gamma _{3}$
is affected in this case, too.

\subsection{General Y-junction}

We start as described above in Sec.\ \ref{sec:perturb1}, but have to
slightly modify our approach afterwards. It turns out that in the case of a 
general Y-junction the intermediate expressions produced in computer
calculation show enormous complexity in the third order. The formulation in
terms of the $S$-matrix followed in the above analysis becomes impractical.

However, we may find the logarithmic corrections to the conductance
directly, in the spirit of our treatment of the impurity in the Luttinger
liquid. \cite{Aristov2009} Indeed, ultimately we need to analyze only the
partial contributions to the two conductances, Eq.\ (\ref{def:ab}), rather
than the full corrections to the $S$-matrix, i.e.\ $3\times 3$ matrix
quantities. Thus we may keep only these two partial contributions from each
diagram, which drastically simplifies the calculation.

Another simplification occurring in the present analysis of the d.c. limit is the
absence of vertex corrections in the diagrams for the conductances. 
\cite{Aristov2009} As a result, it suffices to consider only the self-energy
parts of the Green's functions, corresponding to the above Eq.\ (\ref{Smatexpa}). 
The absence of vertex corrections is easily proven in case of $T=0$ 
which we consider in this paper. The analysis at $T\neq0$ which was undertaken 
for the impurity in the Luttinger liquid, \cite{Aristov2009} is more involved in case of 
the Y-junction and shall be given elsewhere.

Below we explain a few points of this analysis in more detail. Each diagram
contributing, e.g., to the component $G_{a}=(1-a)/2$ in (\ref{def:ab}), stems
from the generic expression $a=\tfrac{1}{2}Tr(\lambda _{3}G_{\omega
=+0}^{R}\lambda _{3}G_{\omega =+0}^{A})$. Here we introduced the retarded
(advanced) Green's functions $G^{R(A)}=\mp i\vartheta (\pm t)\langle \lbrack
\psi (x,t),\psi ^{\dagger }(y,0)]\rangle $ with $x\rightarrow -\infty $, $%
y\rightarrow \infty $. When Fourier-transformed and taken at zero energy,
the Green's functions are stripped from the coordinate dependence and become
simply proportional to the $S$-matrix, $G^{R}=-iS^{\dagger }$, $G^{A}=iS$,
(we set $v_{F}=1$). The expansion (\ref{Smatexpa}) leads to the renormalized
quantity 
\begin{equation}
\begin{aligned} a_{r} & = a_{0} + g a_{1}  + g^{2} a_{2}+ g^{3} a_{3}  +
\ldots \end{aligned}
\end{equation}%
where $a_{0}=\tfrac{1}{2}Tr(\tilde{\lambda}_{3}\lambda _{3})$ and 
\begin{eqnarray}
a_{1} &=& \tfrac12 Tr\left[\tilde \lambda_{3} ( \Sigma_{1}
\lambda_{3} + \lambda_{3} \Sigma_{1}^{\dagger} ) \right] ,
\nonumber \\ 
a_{2} &=&
\tfrac12 Tr\left[\tilde \lambda_{3} ( \Sigma_{2} \lambda_{3} + \Sigma_{1}
\lambda_{3} \Sigma_{1}^{\dagger} + \lambda_{3} \Sigma_{2}^{\dagger} )
\right], \\ 
a_{3} &=& \tfrac12 Tr\left[\tilde \lambda_{3} (\Sigma_{3}
\lambda_{3} + \Sigma_{2} \lambda_{3} \Sigma_{1}^{\dagger} + \Sigma_{1}
\lambda_{3} \Sigma_{2}^{\dagger} + \lambda_{3} \Sigma_{3}^{\dagger} )
\right] ,
\nonumber  
\end{eqnarray}%
and similarly for $b_{r}$, where the components $\Sigma _{j}$ of the self
energy have been defined in (\ref{Smatexpa}). The advantage of directly
calculating corrections to $a_{r},b_{r}$ is most apparent when dealing with
the long expressions for $\Sigma _{3}$. The partial contribution to $a_{3}$
has the form 
\[
\begin{aligned}
&\tfrac{1}{2}Tr\left[ \tilde{\lambda}_{3}(\Sigma _{3}\lambda _{3}+\lambda
_{3}\Sigma _{3}^{\dagger })\right] =\tfrac{1}{2}Tr\left[ \tilde{\lambda}%
_{3}\Sigma _{3}\lambda _{3}\right] +h.c. \\
&=Re(Tr(\lambda _{3}\tilde{\lambda}%
_{3}\Sigma _{3})).
\end{aligned}
\]

The expressions $a_{r}$, $b_{r}$ obtained in this way show both, scale-dependent terms in the form of the logarithm of $\Lambda=L/a$ and in addition scale-independent
terms, as is explained at length elsewhere. \cite{Aristov2009} The result of
this calculation provides expressions for the renormalized values of
conductance, $a_{r},b_{r}$, in terms of the bare ones, $a,b$. In the next
step of the Callan-Symanzik scheme we are considering here, we invert these
expansions, and express the bare values via the renormalized ones, keeping
terms up to $g^{3}$ in the corresponding series.

Then we require that the bare values be independent of the scale in
consideration, i.e. the logarithm $\Lambda $, 
\begin{equation}
\begin{aligned} \frac{d a}{d\Lambda} & = \frac{\partial a}{\partial \Lambda}
+ \frac{\partial a}{\partial a_{r}} \frac{\partial a_{r}}{\partial \Lambda}
+ \frac{\partial a}{\partial b_{r}} \frac{\partial b_{r}}{\partial \Lambda}
=0 \end{aligned}
\end{equation}%
and similarly for $db/d\Lambda $. Solving this system of linear equations
for the quantities $da_{r}/d\Lambda $, $db_{r}/d\Lambda $, we find:

\begin{equation}
\frac{d}{d\Lambda} \begin{pmatrix} a_{r} \\ b_{r} \end{pmatrix} =- %
\begin{pmatrix} {\partial a}/ {\partial a_{r}} , & {\partial a}/{\partial
b_{r}} \\ {\partial b}/ {\partial a_{r}} , & {\partial b}/{\partial b_{r}}
\end{pmatrix} ^{-1} \frac{d}{d\Lambda} \begin{pmatrix} a \\ b \end{pmatrix}
\label{RG-CS}
\end{equation}
This calculation is best done by means of computer algebra, since the
intermediate expressions are quite complicated.

We keep the terms of the order of $g^{3}$ in the final expressions. A
criterion for the correctness of the calculation is the absence of any $%
\Lambda $-dependence on the right-hand side of Eq.\ (\ref{RG-CS}) up to this
order.

When we compare the final expressions, found in this direct calculation,
with the first terms of the Taylor expansion of the corresponding
expressions (\ref{difeq3}), (\ref{def:alphas}), we find complete agreement
of the universal (regularization independent) parts of the $\beta $-functions to
third order. In addition to these universal contributions of the
``one-loop''  ladder summation (\ref{def:alphas}), 
we also find non-ladder contributions $\beta _{3}$, which
first appear in third order and explicitly contain the above mentioned
regularization-dependent coefficients $b_{j}$. These contributions are
rather complicated and are listed in Appendix.

Comparing these three-loop contributions with the simpler case of one impurity in the Luttinger liquid, we should make several remarks. First of all, one can explicitly check 
that these contributions satisfy Eq.\ (\ref{GperpRG}) so that the curve $G_{\perp}=1$ remains the fixed line. Second, for small $g,g_{3}$ these terms do not lead to the appearance of extra FPs, but the situation with strong interaction $|g|,|g_{3}| \sim 1$ is, strictly speaking, unclear. However, we make a plausible conjecture, that all FPs of full RG beta function (which contains three loop, four loop etc.\ contributions, beyond the re-summed one-loop terms (\ref{difeq4}), (\ref{def:alphas}) ) lie on the borderline of allowed conductances for the non-interacting case. 

The next remark concerns the universality of the phase diagram, proposed below in Sec.\ref{sec:discussion} 
on the basis of the expression  (\ref{difeq4}), (\ref{def:alphas}). What is the evidence that no new FPs appear at strong interaction, and the structure of the phase diagram is independent of regularization ? The answer to this question goes along several lines.
From the actual form of three-loop contributions we see that regularization does not change the position of the interaction-dependent FP $M$ only in fully symmetric case. 
However, in the latter case the three loop terms are apparently unimportant even in the strongly interacting regime, as is suggested by comparison of our results with those in \cite{Oshikawa2006}. The bosonization approach by Oshikawa et al. showed changes in the character of $A$ fixed point at strong attraction, $K=K_{3} = 3$, and we see the appearance of FP $Q$ at this value. The scaling dimensions of leading perturbations, found in Sec.\  10.1 and 10.4.2 of \cite{Oshikawa2006}, agree with our one loop formulas (\ref{RGeq:N}), (\ref{RGeq:A}) for the whole range of interaction strength.  

In summary, we again find that the non-universal terms in the $\beta $-functions do not influence the behavior at the fixed points $N,A$, which means that the ladder summation is sufficient in the case of repulsive interactions. As for the fixed point $M$, of importance for attractive interaction, we find that the non-universal terms are relevant and may change the power law exponents in principle. We address this question in some detail in the next subsection. 

\subsection{\label{sec:FulSym3loop}Fully symmetric Y junction}

In the important case of equal interactions, $g_{3}=g$, the position of the
third fixed point ($M$) in Eq.\ (\ref{explRGladder}) is $\cos \theta =x_{3}=-1/3$
and remains unaffected by $\beta _{3}$ in (\ref{RG_beta3}). 
Expanding $\beta _{3}$ around $x_{3}$ we have 
\begin{equation}
\beta _{3}=\frac{\sqrt{2}}{18}g^{3}(x+1/3)(4b_{2}-b_{3}+b_{4})+\ldots ,
\label{beta3x}
\end{equation}%
so that the scaling exponent along the limiting parabola, $\gamma _{3}$ in Eq.\ (\ref{exponents-sym}), depends on the
non-universal coefficients $b_{j}$, which were found to depend on the
regularization of the theory. The situation is however more delicate, because the scaling exponent $\gamma'_{3} $ in (\ref{expMpoint}) in the direction perpendicular to parabola remains unchanged by the three loop contributions. To see that we expand the expressions for $\beta_{3}$ listed in Appendix \ref{sec:AppB3}. In terms of variables $\bar a$, $c$
we have the additional three loop contributions to Eq.\   (\ref{difeqFullsym}) 
\begin{equation}
\begin{aligned} 
\frac{d \bar a }{d\Lambda} =&  -g^{3} \frac{ 2b_{2}+b_{4}}{16}  
(1-\bar a)^{2} (1+3\bar a)^{2}
+  {\cal O}(g^{3}c^{2}) \,, \\ 
\frac{d c}{d\Lambda} =& g^{3} c \frac{ \bar a(1-\bar a)}{8}  
\left( 2b_{2}(5+7 \bar a)+b_{3}(\bar a-1) 
\right. \\ & \left. 
+4b_{4} (1+2\bar a)
\right) 
+ {\cal O}(g^{3} c^{2}) \,.
\end{aligned}
\label{3loopFullsym}
\end{equation}
The second equation here at the $M$ fixed point $\bar a=-1/3$ corresponds to (\ref{beta3x}). The first equation (\ref{3loopFullsym}) together with (\ref{difeqFullsym}) show that near the fixed points the  three loop contributions do not change the scaling exponent  $\gamma'_{3} $ of Eq.\ (\ref{expMpoint}). 

\section{\label{sec:discussion}Discussion and conclusions}

In this paper we presented a theory of charge transport through a junction of three quantum wires, modelled by Luttinger liquids. We focused on the case of a Y-junction, a set-up symmetric with respect to interchanging wires $1$ and $2$. We allow for different interaction strengths $g$ and $g_{3}$ in wires $1,2$ and wire $3$, respectively. Our method employs a purely fermionic representation, which has the advantage that the connection to ideal (noninteracting) leads is naturally incorporated. The transition from the noninteracting leads to the interacting wire is assumed to be adiabatic. We find that at zero temperature the scattering process is completely described by elastic 
scattering (no excitation of real particle-hole pairs). Virtual excitations of multi particle-hole pairs are all-important; these processes are described in terms of the renormalized single particle S-matrix. In terms of diagrams for the conductances this amounts to the absence of any vertex corrections (at $T=0$). 

We extended a theory previously applied to a two-wire junction to the Y-junction problem. That theory employs perturbation theory with respect to the interaction in fermionic language (using, however, the concepts of current algebra to systematize the bookkeeping) to derive the renormalization group $\beta$-function for the conductance. As shown by us in Ref.\ \cite{Aristov2009}  an RPA type ladder summation of an infinite class of terms of perturbation theory may be performed to generate all of the known results on the scaling behavior of the conductance, power law exponents, crossover behavior, and more, for any interaction strength and any scattering characteristic of the barrier. 

In the general time-invariant case the tensor of conductances features two independent components $G_{a}$, $G_{b}$. These components are confined to an area bounded by a curve $B$ in the fundamental domain $0<G_{a},G_{b}<1$. We derive the coupled set of RG-equations for conductances in the ladder approximation. It is interesting to note that the fixed points of these equations are all located on the 
boundary curve $B$. In fact a simplified tunneling model leads to conductances located on the boundary curve $B$, and allows for an analytical determination of the fixed points and the conductances. 

We probe the validity of the ladder summation by evaluating all contributions up to and including third order (several thousands diagrams). We classify the contributions into universal (ladder summation) and non-universal, with respect to the regularization (finite length $L$  or finite temperature $T$). For repulsive interaction we find that the non-universal contributions to the RG-$\beta$-functions are subleading in the scaling regime, indicating that the ladder summation is fundamentally correct in the vicinity of the stable fixed point. For attractive interaction we find that the non-universal contribution in fact changes the location of the fixed points and the values of the exponents.

\begin{figure}[tbp] 
\includegraphics*[width=\columnwidth]{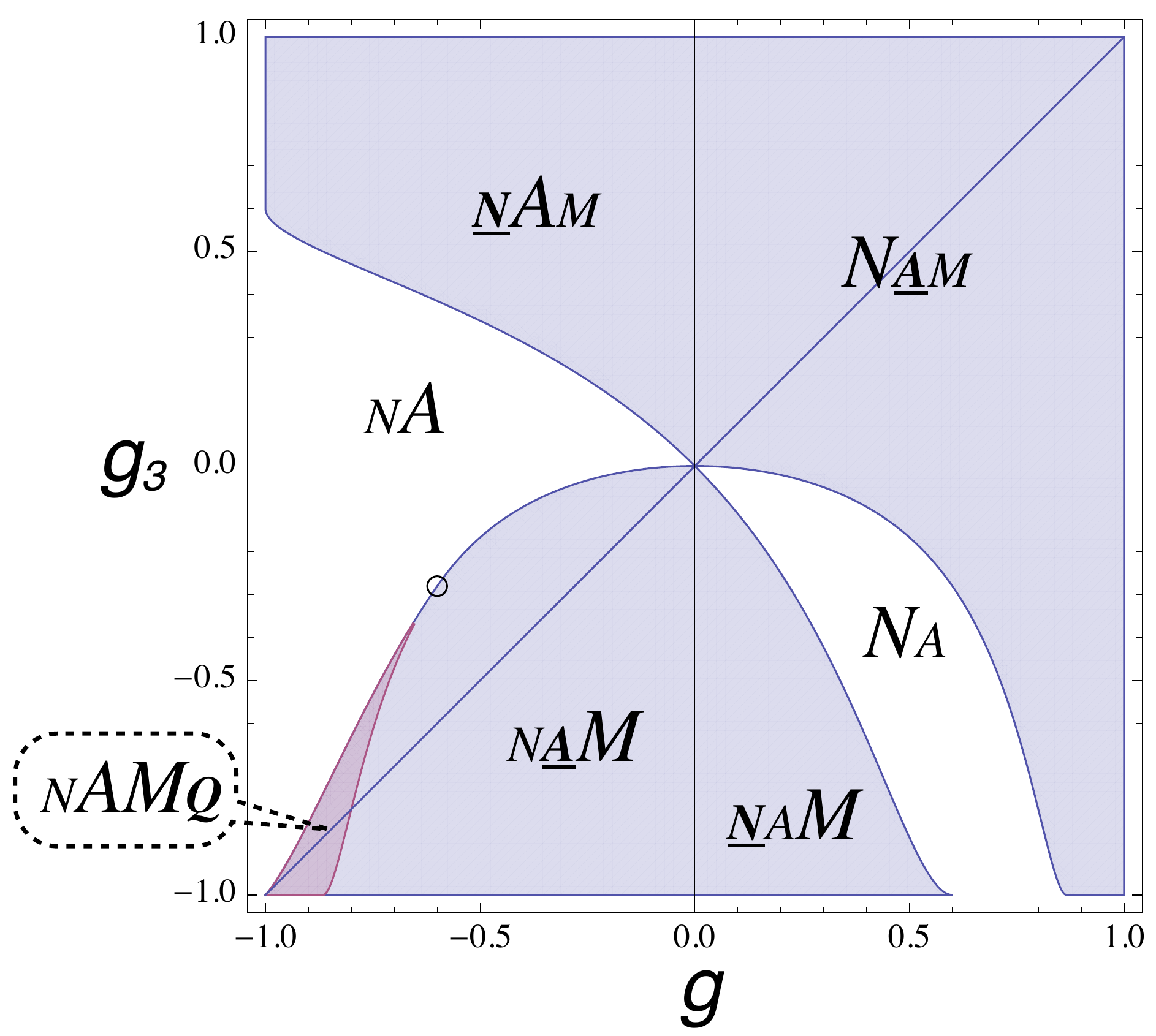}
\caption{(Color online) The number and the type of physically available fixed points is shown for various strength of interactions $g$ and $g_{3}$, as defined by the ladder summation. The positions of N and A points are independent of interaction, the positions of M and Q points are discussed in text. The stable, unstable and saddle-point FPs are denoted by large, small and underlined capital letters, respectively. See text for additional details. }
\label{fig:RGpoints}
\end{figure}

We find a rich scenario of fixed points. In total there are four fixed points $N,A,M,Q$, but not all of them are in the physically accessible regime. Figure \ref{fig:RGpoints} shows the distribution of fixed points in the coupling constant $g-g_{3}$-plane. In each regime the stable, unstable and saddle-point (unstable) fixed points are indicated by large, small and underlined capital letters. For instance, in the case of repulsive interactions FP $N$ describes the totally separated wires; it is the stable fixed point . FP $A$ becomes a stable FP for strong attractive interaction. It stands for wire $3$ separated from the ideal wire $1-2$. For any attractive interaction with $g=g_{3}$ the FP $M$ is stable. It corresponds to finite conductances in all ways, the value depending on the interaction strengths $g,g_{3}$. We conjecture that $M$ corresponds to the "mystery point" discussed in \cite{Oshikawa2006} for the totally symmetric junction threaded by magnetic flux. The FP $Q$, finally, is always unstable, of the saddle point  character, in the limited region where it enters the physical domain.

Most neighboring regions in Fig.\  \ref{fig:RGpoints} differ by one FP $M$, which appears either at the $A$ point or at the $N$ point, with the corresponding change in the character of this latter point. The situation at the interface between {\it{\small N}{\large A}} and 
{\it{\small N}{\large AM}{\small Q}}  is different, as both ``floating'' FPs $M$ and $Q$ appear at one point on the parabola (see Fig.\ \ref{fig:RGflowK37}), away from $A$ or $N$. We recall, that for equal interaction strength, $g=g_{3}$, the $Q$ point appears first at $K=K_{3}=3$.

It might be also interesting to note the existence of the ``tricritical'' point between the phases {\it{\small N\underline{A}}{\large M}}, {\it{\small N}{\large A}} and {\it{\small N}{\large AM}{\small Q}}, which happens at $K=2$, $K_{3}=4/3$, (i.e., at $g=-3/5$, $g_{3}=-7/25$) see Sec.\ \ref{sec:ExplicitLadder}. This tricritical point corresponds to the  situation when both $M$ and $Q$ points merge with the $A$ point. 

Comparing our findings with previous studies, we observe that 
there is a correspondence between the scaling exponents $\gamma_{3}$ in our
Eq.\ (\ref{exponents-sym}) and $(1-K_{3}^{-1})$ for $b'$ in (\ref{RGeq:A}),  and those numerically obtained by fRG method in 
 \cite{Barnabe2005a},  respectively $\gamma_{2}(3)$ in Eq.\ (56) and $\gamma_{1}^{\rm HF}(3)$ in Eq.\ (48) there. 
 
We confirm that the exponents obtained by our method around N point, Eqs.\ (\ref{RGeq:N}),  and A point, (\ref{RGeq:A}),  at $K_{3}=K$, coincide with those obtained by bosonization  in \cite{Oshikawa2006},  Eq.\ (10.23),  and Eqs.\ (10.106), (10.107), respectively.  At the same time, the exponent $(1-K^{-1})$ around the N point both in our work and Ref.\ [\onlinecite{Oshikawa2006}] differs from the exponent $1-(3K)^{-1}$ reported in Eq.\  (2.11) of Ref.\  [\onlinecite{Nayak1999}] ; we note that in the latter case the RG flow exists even in the absence of interaction, $K=1$. It was argued in \cite{Nayak1999,Oshikawa2006} that the bosonic theory of the Y-junction has the duality property $K \to 3/K$, which corresponds to a change from the case of totally separated wires (N point) to the case of maximally open Y-junction (D point). At the latter the conductance exceeds the value allowed by the unitarity of the single-particle S-matrix, as argued there possibly due to the formation of Cooper pairs at strong attraction.  The scaling dimension of the leading perturbation around this hypothetic D point was thus found as  $K/3$, in Eq.\ (10.30) of \cite{Oshikawa2006}, and as 
$K/9$ in Eq.\ (3.3) of \cite{Nayak1999}, the latter value evidently arising due to the above additional factor $1/3$ around the N point. As a result, bosonization studies predict a qualitative change in the scaling behavior of the system at $K>3$,  \cite{Oshikawa2006}  or at $K>9$. \cite{Nayak1999} Our analysis also shows a qualitative change at $K=K_{3} >3$, which corresponds to the appearance of extra FPs, but not of the exotic D type. We stress again, that the RG flows in our study always end at the surface of the (generally four-dimensional) body describing the conductance matrix in the absence of interaction. Particularly, the D point (which is $a=b=-1$ or $G_{a}=1$, $G_{b}=4/3$ in our notation, see \cite{Aristov2011}) is not  a fixed point of Eqs.\ (\ref{difeq3}) and the RG trajectories do not end there, even if we start from outside this body. Our above analysis of the three loop RG contributions confirms this picture. We have looked for contributions violating the unitarity condition as proposed in \cite{Oshikawa2006}, but did not find any. In our formalism such contributions would be generated by vertex corrections. However, at $T=0$, all vertex corrections vanish in the d.c. limit. 

In summary, we have derived a renormalization group theory description for the two independent conductances characterizing a Y-junction of a Luttinger liquid wire (1,2) with interaction constant $K$ and a tunneling tip Luttinger liquid wire with interaction constant $K_{3}$. We summed up infinite classes of contributions in perturbation theory (ladder approximation) to obtain the RG $\beta$-functions. Additional contributions appearing in third order were employed to decide whether the result of the ladder approximation in the neighborhood of the stable fixed points remained unchanged. This was found to be the case for repulsive interaction (where arguments can be made that all higher order non-ladder terms should also be negligible). In the case of attractive interaction non-ladder contributions might change the critical behavior in certain cases. The existence of further fixed points, not captured by the ladder approximation, cannot be excluded, although it is not very likely. Nonetheless, we find it remarkable that our method allows to determine a rich scenario of fixed points and RG-flows, including the crossover behavior. The corresponding conductances as a function of the scaling parameter are readily accessible.  


\begin{acknowledgments}
We thank D.A. Bagrets,  A.P. Dmitriev, V.Yu. Kachorovskii, A.W.W. Ludwig,  
 I. Safi,  A.G. Yashenkin for useful discussions. 
We are grateful to I.V. Gornyi and D.G. Polyakov for careful reading of the manuscript 
and valuable suggestions. 
 The work of D.A. was supported 
by the German-Israeli Foundation, the DFG-Center for Functional Nanostructures, 
and the Dynasty foundation, RFBR grant 09-02-00229.   
The work of P.W. was supported 
by the DFG-Center for Functional Nanostructures.
\end{acknowledgments}

\appendix

\section{\label{sec:GellMann} Generators of SU(3) group}
For reader's convenience, we list here the matrices $\lambda _{j}$ 
used in the main part of the paper. 
The traceless Gell-Mann matrices, $\lambda _{j}$, with $j=1,\ldots 8$
are the generators of the SU(3) group, discussed, e.g., in  
\cite{Tilma2002}.  
\begin{equation}
\begin{aligned}
\lambda _{1}&=\begin{pmatrix} 0 & 1 & 0 \\
 1 & 0 & 0 \\
 0 & 0 & 0 
\end{pmatrix} , \quad 
\lambda _{2}=\begin{pmatrix}  0 & -i & 0 \\
 i & 0 & 0 \\
 0 & 0 & 0 \end{pmatrix}  ,  \\
\lambda _{3}&=\begin{pmatrix} 1 & 0 & 0 \\
 0 & -1 & 0 \\
 0 & 0 & 0 \end{pmatrix}   , \quad
\lambda _{4}=\begin{pmatrix}  0 & 0 & 1 \\
 0 & 0 & 0 \\
 1 & 0 & 0 \end{pmatrix}  ,  \\
\lambda _{5}&=\begin{pmatrix}  0 & 0 & -i \\
 0 & 0 & 0 \\
 i & 0 & 0 \end{pmatrix}  ,  \quad
\lambda _{6}=\begin{pmatrix}  0 & 0 & 0 \\
 0 & 0 & 1 \\
 0 & 1 & 0 \end{pmatrix}  ,  \\
\lambda _{7}&=\begin{pmatrix}  0 & 0 & 0 \\
 0 & 0 & -i \\
 0 & i & 0 \end{pmatrix}  ,  \quad
\lambda _{8}=\frac{1}{\sqrt{3}}\begin{pmatrix} 1 & 0 &0\\ 0 & 1 &0\\ 0 & 0 &-2
\end{pmatrix}   
\end{aligned}
\end{equation}
Together with the unit matrix $\lambda _{0}$,  they have the property $Tr[\lambda_{j}\lambda _{k}]=2\delta _{jk}$, with $j,k=0,\ldots 8$. After this normalization, the structure of algebra is determined by the structure constants $f_{ijk}$ according to $[\lambda_{j}, \lambda _{k}] = 2 i \sum_{l}f_{jkl} \lambda_{l}$. In the familiar case of SU(2) algebra (which is a subalgebra of SU(3), spanned by $\lambda _{1,2,3}$),  one has $f_{jkl} = \epsilon_{jkl}$, with totally antisymmetric tensor; it leads to simple mnemonic rules.  For the present SU(3) case such simple rules are absent, and in most cases we resorted to symbolic computer calculations, which are readily done, e.g., in  \emph{Mathematica}.

\section{\label{sec:AppB3} $\protect\beta_{3}$ for the general Y-junction}

The perturbative calculation of the corrections to the conductances, as
described above, eventually leads to RG $\beta $%
-functions containing one-loop and three-loop contributions. The one-loop
contributions are resummed into expressions (\ref{def:alphas}), and the
three-loop contributions to $da/d\Lambda $, $db/d\Lambda $ are found as
functions of $a,b$, Eq.\ (\ref{def:ab}). 
\begin{equation}
\begin{aligned} \beta_{3a} &=  g^{3} (F_{2}b_{2} +
F_{3}b_{3} + F_{4}b_{4}) , \\ 
\beta_{3b} &=  g^{3}(F_{2}'b_{2} +
F_{3}'b_{3} + F_{4}'b_{4}) \end{aligned}
\label{3loopab}
\end{equation}
where
\begin{equation}
\begin{aligned}
F_{2} = &\frac{1}{864} \left(-432 a^4-36 a^3 (b-1)
  \kappa +a^2 \left(-24 b^2 \kappa ^2
 \right.\right.  \\  &\left.\left.   +b
  \left(39 \kappa ^2+405\right)-15 \kappa
  ^2+459\right)-4 a \kappa  \left(2 b^3
  \kappa ^2
 \right.\right.  \\  &\left.\left. 
  -3 b^2 \kappa ^2-9 b+\kappa
  ^2+9\right)+3 \left(4 b^3 \kappa ^2-b^2
  \left(5 \kappa ^2
    \right.\right.\right.  \\  &\left.\left.\left.
    +27\right)
   +b
  \left(\kappa
  ^2-81\right)-36\right)\right) , \\
     F_{3} =&  \frac{ 1-b}{288} \left(a^2-b\right)
  \left(6 (a+1) \kappa -(2 b+1) \kappa
  ^2-9\right) ,\\
      F_{4} =&  \frac{1}{432} \left(-108 a^4-18 a^3 (b-1)
  \kappa -3 a^2 \left(b^2 \kappa ^2
        \right.\right.  \\  &\left.\left.
  -2 b   \left(\kappa ^2  
        +18\right) +\kappa
  ^2-36\right)-a \kappa  \left(2 b^3
  \kappa ^2
           \right.\right.  \\  &\left.\left.
  -3 b^2 \left(\kappa
  ^2+3\right)+\kappa ^2+9\right)-27
  (b+1)^2\right) .
  \end{aligned}
\end{equation}
  \begin{equation}
\begin{aligned}
F_{2}' =&  -\frac{1}{864} (b-1) \left(216 a^3+72 a^2
  (b-1) \kappa 
  \right.  \\  &\left.
  +3 a \left(4 b^2 \kappa
  ^2+b \left(\kappa ^2-45\right)-5 \kappa
  ^2-27\right)
           \right. \\  &\left.
  +4 (b-1) (2 b+1)^2 \kappa
  ^3\right) ,\\
     F_{3}' =&  \frac{1}{288} a (b-1)^2 \left(6 (a+1)
  \kappa -(2 b+1) \kappa ^2-9\right) ,\\
   F_{4}' =&   \frac{1-b}{432} \left(54 a^3+9 a^2
  (b-1) \kappa
   +3 a \left(2 b^2 \kappa   ^2
     \right.\right.  \\  &\left.\left.
  -b \left(\kappa ^2+9\right)-\kappa
  ^2-9\right)+(b-1) (2 b+1)^2 \kappa
  ^3\right) ,      
\end{aligned}
\end{equation}
and $\kappa =1+2g_{3}/g$.
These expressions are used for the analysis of the fixed point $M$ in Sec.\ \ref{sec:FulSym3loop}. 
In order to compare to the result
of the simple tunneling case, Eq.\ (\ref{RG_beta3}), and to our previous
work, it is more convenient to go now from the "conductances" $a,b$ to the
angular quantities $\theta ,\psi $ and discuss three-loop contributions to $%
d\theta /d\Lambda $ and $d\psi /d\Lambda ,$ which we denote by $\beta
_{3\theta }$ and $\beta _{3\psi }$, respectively.

\begin{equation}
\begin{aligned} \beta_{3\theta} &= g^{3}\sin \theta \sin^{2} \psi \cos \psi
\frac{2b_{2}+b_{4}}{16} \\ &+ \frac{g^{3}}{128}\sin^{3}\theta (f_{2}b_{2} +
f_{3}b_{3} + f_{4}b_{4}) , 
\end{aligned}
\end{equation}\begin{equation}
\begin{aligned}
\\ 
\beta_{3\psi} &= g^{3}\sin^{3}\psi \cos\theta \frac{2b_{2} +
b_{4}}{4} \\ & + \frac{g^{3}}{128} \sin\psi \sin^{2} \theta (f_{2}'b_{2} +
f_{3}'b_{3} + f_{4}'b_{4}) \end{aligned}
\label{3loop}
\end{equation}%
with the functions 
\begin{equation}
\begin{aligned}  f_{2} &= 4\kappa \cos \theta (2 \cos^{2} \psi
+\kappa^{2} \cos^{2} \theta) \\ & + \cos\psi (8\cos 2 \psi -7 +
\kappa^{2}(2+\cos 2 \theta)) \\ f_{3}&= \cos\psi (3 - 2 \kappa (1+\cos\theta
\cos\psi)+\kappa^{2}\cos^{2}\theta )\\ f_{4}& = 2\kappa\cos\theta
(\cos^{2}\psi + \kappa^{2}\cos^{2}\theta \\ & + \kappa\cos\theta \cos \psi )
+2 \cos 3 \psi \end{aligned}
\end{equation}%
and 
\begin{equation}
\begin{aligned}   f_{2}' &= \frac{3 \cos^{2} \theta
-1}{\cos\theta} ( \kappa^{2} \cos 2 \theta - 1) \\ & + 8 \kappa \cos\psi +
24 \cos\theta \cos 2\psi \\ f_{3}'&= \frac{1-3 \cos^{2} \theta }{\cos\theta}
(3- 2\kappa + \kappa^{2} \cos^{2}\theta \\& - 2 \kappa \cos\theta \cos\psi)
\\ f_{4}'& = 2\frac{1- 3 \cos^{2} \theta }{\cos\theta} + 12 \cos \theta \cos
2\psi \\ & + 2 \kappa (1+3\cos^{2}\theta ) \cos \psi \end{aligned}
\end{equation}%
The previously considered case of impurity in the 
Luttinger liquid \cite{Aristov2009} corresponds to setting $\theta=0$ in the above equations. From the form of (\ref{3loop}) and Eq.\ (\ref{difeq2}) one verifies that the scaling exponents at the RG fixed points $N$ ($\theta=0$, $\psi=0$) and $A$ ($\theta=\pi$, $\psi=0$) are not modified by the presence of  $\beta_{3\theta}$,  $\beta_{3\psi}$, as the latter functions contain higher powers of $\theta$, $\psi$ in those points. 

\section{Fixed point  $M$ in asymmetric case}

In this section we consider the position of the FP $M$
in case when all three bulk interaction terms $g_{j}$ in Eq.\ (\ref{iniHam}) are different. 
We restrict ourself by the first order in $g_{j}$, and 
use the precursor to the RG equation (\ref{RG:dS}) in its general form 
\begin{equation}
S^{-1}.S_{r}-1=- \tfrac 12\Lambda \sum_{j=1}^{3} g_{j} [\tilde \rho_{j}, \rho_{j}] \,.
\label{RG:asym}
\end{equation}
We parametrize \cite{Aristov2011} the $S$-matrix in $T$-symmetric case by 
\begin{equation}
\begin{aligned}
S & = e^{i \lambda_{2} \xi/2}e^{i \lambda_{3} (\pi-\psi)/2} 
e^{i \lambda_{5} \theta} e^{i \lambda_{2} \xi/2} \,.
\end{aligned}
\label{Euler2}
\end{equation}
The conductance matrix in (\ref{defGprime}) is then given by 
\begin{eqnarray} 
G_{aa} & =&
\tfrac12 \sin^{2}\xi \left(1- \cos \theta \cos\psi  \right)
+ \tfrac 14 \cos^{2}\xi  \sin^{2} \theta  \,, \nonumber\\ 
G_{ab} & =& G_{ba}  = \cos \xi  \sin^{2}\theta \,,  \quad 
G_{bb}  = \sin^{2}\theta \,.
\end{eqnarray}
so that the Eq.\  (\ref{Gmatrixhat}) is restored at $\xi=\pi/2$. 

Simple calculations show that r.h.s.\ of (\ref{RG:asym}) is zero at $N$ fixed point, $\theta=0$, $\psi=0$, and at three $A$ points ~: i) $\theta=\pi$, $\psi=0$, $\xi=\pi/2$, considered above, ii) $\theta= \pi/2$, $\xi=0$,  iii) $\theta= \pi/2$, $\xi=\pi$. The position of $M$ point is defined by conditions $\psi=0$ and 
\begin{equation}
\begin{aligned}
\cos \theta & =-g_{3}^{-1} /( g_{1}^{-1}+g_{2}^{-1}+g_{3}^{-1})\,, \\ 
\cos \xi & =   (g_{2}^{-1}-g_{1}^{-1}) /( g_{1}^{-1}+g_{2}^{-1}+2g_{3}^{-1}) \, , 
\end{aligned}
\label{genMpoint}
\end{equation}
at $g_{1}=g_{2}=g$ we return to the formulas in main part of the paper, $\psi=0$, $\xi=\pi/2$, and $\cos\theta$ defined by Eq.\ (\ref{RGeq-prec}).


\begin{thebibliography}{30}
\expandafter\ifx\csname natexlab\endcsname\relax\def\natexlab#1{#1}\fi
\expandafter\ifx\csname bibnamefont\endcsname\relax
 \def\bibnamefont#1{#1}\fi
\expandafter\ifx\csname bibfnamefont\endcsname\relax
 \def\bibfnamefont#1{#1}\fi
\expandafter\ifx\csname citenamefont\endcsname\relax
 \def\citenamefont#1{#1}\fi
\expandafter\ifx\csname url\endcsname\relax
 \def\url#1{\texttt{#1}}\fi
\expandafter\ifx\csname urlprefix\endcsname\relax\def\urlprefix{URL }\fi
\providecommand{\bibinfo}[2]{#2}
\providecommand{\eprint}[2][]{\url{#2}}

\bibitem[{\citenamefont{Bockrath et~al.}(1999)\citenamefont{Bockrath, Cobden,
  Lu, Rinzler, Smalley, Balents, and McEuen}}]{Bockrath1999}
\bibinfo{author}{\bibfnamefont{M.}~\bibnamefont{Bockrath}},
  \bibinfo{author}{\bibfnamefont{D.~H.} \bibnamefont{Cobden}},
  \bibinfo{author}{\bibfnamefont{J.}~\bibnamefont{Lu}},
  \bibinfo{author}{\bibfnamefont{A.~G.} \bibnamefont{Rinzler}},
  \bibinfo{author}{\bibfnamefont{R.~E.} \bibnamefont{Smalley}},
  \bibinfo{author}{\bibfnamefont{L.}~\bibnamefont{Balents}}, \bibnamefont{and}
  \bibinfo{author}{\bibfnamefont{P.~L.} \bibnamefont{McEuen}},
  \bibinfo{journal}{Nature} \textbf{\bibinfo{volume}{397}},
  \bibinfo{pages}{598} (\bibinfo{year}{1999}).

\bibitem[{\citenamefont{Yao et~al.}(1999)\citenamefont{Yao, Postma, Balents,
  and Dekker}}]{Yao1999}
\bibinfo{author}{\bibfnamefont{Z.}~\bibnamefont{Yao}},
  \bibinfo{author}{\bibfnamefont{H.~W.~C.} \bibnamefont{Postma}},
  \bibinfo{author}{\bibfnamefont{L.}~\bibnamefont{Balents}}, \bibnamefont{and}
  \bibinfo{author}{\bibfnamefont{C.}~\bibnamefont{Dekker}},
  \bibinfo{journal}{Nature} \textbf{\bibinfo{volume}{402}},
  \bibinfo{pages}{273} (\bibinfo{year}{1999}).

\bibitem[{\citenamefont{Nayak et~al.}(1999)\citenamefont{Nayak, Fisher, Ludwig,
  and Lin}}]{Nayak1999}
\bibinfo{author}{\bibfnamefont{C.}~\bibnamefont{Nayak}},
  \bibinfo{author}{\bibfnamefont{M.~P.~A.} \bibnamefont{Fisher}},
  \bibinfo{author}{\bibfnamefont{A.~W.~W.} \bibnamefont{Ludwig}},
  \bibnamefont{and} \bibinfo{author}{\bibfnamefont{H.~H.} \bibnamefont{Lin}},
  \bibinfo{journal}{Phys. Rev. B} \textbf{\bibinfo{volume}{59}},
  \bibinfo{pages}{15694} (\bibinfo{year}{1999}).

\bibitem[{\citenamefont{Safi et~al.}(2001)\citenamefont{Safi, Devillard, and
  Martin}}]{Safi2001}
\bibinfo{author}{\bibfnamefont{I.}~\bibnamefont{Safi}},
  \bibinfo{author}{\bibfnamefont{P.}~\bibnamefont{Devillard}},
  \bibnamefont{and} \bibinfo{author}{\bibfnamefont{T.}~\bibnamefont{Martin}},
  \bibinfo{journal}{Phys. Rev. Lett.} \textbf{\bibinfo{volume}{86}},
  \bibinfo{pages}{4628} (\bibinfo{year}{2001}).

\bibitem[{\citenamefont{Yi}(2002)}]{Yi2002}
\bibinfo{author}{\bibfnamefont{H.}~\bibnamefont{Yi}}, \bibinfo{journal}{Phys.
  Rev. B} \textbf{\bibinfo{volume}{65}}, \bibinfo{pages}{195101}
  (\bibinfo{year}{2002}).

\bibitem[{\citenamefont{Lal et~al.}(2002)\citenamefont{Lal, Rao, and
  Sen}}]{Lal2002}
\bibinfo{author}{\bibfnamefont{S.}~\bibnamefont{Lal}},
  \bibinfo{author}{\bibfnamefont{S.}~\bibnamefont{Rao}}, \bibnamefont{and}
  \bibinfo{author}{\bibfnamefont{D.}~\bibnamefont{Sen}},
  \bibinfo{journal}{Phys. Rev. B} \textbf{\bibinfo{volume}{66}},
  \bibinfo{pages}{165327} (\bibinfo{year}{2002}).

\bibitem[{\citenamefont{Chen et~al.}(2002)\citenamefont{Chen, Trauzettel, and
  Egger}}]{Chen2002}
\bibinfo{author}{\bibfnamefont{S.}~\bibnamefont{Chen}},
  \bibinfo{author}{\bibfnamefont{B.}~\bibnamefont{Trauzettel}},
  \bibnamefont{and} \bibinfo{author}{\bibfnamefont{R.}~\bibnamefont{Egger}},
  \bibinfo{journal}{Phys. Rev. Lett.} \textbf{\bibinfo{volume}{89}},
  \bibinfo{pages}{226404} (\bibinfo{year}{2002}).

\bibitem[{\citenamefont{Moore and Wen}(2002)}]{Moore2002}
\bibinfo{author}{\bibfnamefont{J.~E.} \bibnamefont{Moore}} \bibnamefont{and}
  \bibinfo{author}{\bibfnamefont{X.-G.} \bibnamefont{Wen}},
  \bibinfo{journal}{Phys. Rev. B} \textbf{\bibinfo{volume}{66}},
  \bibinfo{pages}{115305} (\bibinfo{year}{2002}).

\bibitem[{\citenamefont{Chamon et~al.}(2003)\citenamefont{Chamon, Oshikawa, and
  Affleck}}]{Chamon2003}
\bibinfo{author}{\bibfnamefont{C.}~\bibnamefont{Chamon}},
  \bibinfo{author}{\bibfnamefont{M.}~\bibnamefont{Oshikawa}}, \bibnamefont{and}
  \bibinfo{author}{\bibfnamefont{I.}~\bibnamefont{Affleck}},
  \bibinfo{journal}{Phys. Rev. Lett.} \textbf{\bibinfo{volume}{91}},
  \bibinfo{pages}{206403} (\bibinfo{year}{2003}).

\bibitem[{\citenamefont{Pham et~al.}(2003)\citenamefont{Pham, Pi\'echon, Imura,
  and Lederer}}]{Pham2003}
\bibinfo{author}{\bibfnamefont{K.-V.} \bibnamefont{Pham}},
  \bibinfo{author}{\bibfnamefont{F.}~\bibnamefont{Pi\'echon}},
  \bibinfo{author}{\bibfnamefont{K.-I.} \bibnamefont{Imura}}, \bibnamefont{and}
  \bibinfo{author}{\bibfnamefont{P.}~\bibnamefont{Lederer}},
  \bibinfo{journal}{Phys. Rev. B} \textbf{\bibinfo{volume}{68}},
  \bibinfo{pages}{205110} (\bibinfo{year}{2003}).

\bibitem[{\citenamefont{Egger et~al.}(2003)\citenamefont{Egger, Trauzettel,
  Chen, and Siano}}]{Egger2003}
\bibinfo{author}{\bibfnamefont{R.}~\bibnamefont{Egger}},
  \bibinfo{author}{\bibfnamefont{B.}~\bibnamefont{Trauzettel}},
  \bibinfo{author}{\bibfnamefont{S.}~\bibnamefont{Chen}}, \bibnamefont{and}
  \bibinfo{author}{\bibfnamefont{F.}~\bibnamefont{Siano}},
  \bibinfo{journal}{New Journal of Physics} \textbf{\bibinfo{volume}{5}},
  \bibinfo{pages}{117} (\bibinfo{year}{2003}).

\bibitem[{\citenamefont{Kim et~al.}(2004)\citenamefont{Kim, Vishveshwara, and
  Fradkin}}]{Kim2004}
\bibinfo{author}{\bibfnamefont{E.-A.} \bibnamefont{Kim}},
  \bibinfo{author}{\bibfnamefont{S.}~\bibnamefont{Vishveshwara}},
  \bibnamefont{and} \bibinfo{author}{\bibfnamefont{E.}~\bibnamefont{Fradkin}},
  \bibinfo{journal}{Phys. Rev. Lett.} \textbf{\bibinfo{volume}{93}},
  \bibinfo{pages}{266803} (\bibinfo{year}{2004}).

\bibitem[{\citenamefont{Rao and Sen}(2004)}]{Rao2004}
\bibinfo{author}{\bibfnamefont{S.}~\bibnamefont{Rao}} \bibnamefont{and}
  \bibinfo{author}{\bibfnamefont{D.}~\bibnamefont{Sen}},
  \bibinfo{journal}{Phys. Rev. B} \textbf{\bibinfo{volume}{70}},
  \bibinfo{pages}{195115} (\bibinfo{year}{2004}).

\bibitem[{\citenamefont{Kazymyrenko and Dou\ifmmode~\mbox{\c{c}}\else
  \c{c}\fi{}ot}(2005)}]{Kazymyrenko2005}
\bibinfo{author}{\bibfnamefont{K.}~\bibnamefont{Kazymyrenko}} \bibnamefont{and}
  \bibinfo{author}{\bibfnamefont{B.}~\bibnamefont{Dou\ifmmode~\mbox{\c{c}}\else
  \c{c}\fi{}ot}}, \bibinfo{journal}{Phys. Rev. B}
  \textbf{\bibinfo{volume}{71}}, \bibinfo{pages}{075110}
  (\bibinfo{year}{2005}).

\bibitem[{\citenamefont{Enss et~al.}(2005)\citenamefont{Enss, Meden,
  Andergassen, Barnab\'e-Th\'eriault, Metzner, and Sch\"onhammer}}]{Enss2005}
\bibinfo{author}{\bibfnamefont{T.}~\bibnamefont{Enss}},
  \bibinfo{author}{\bibfnamefont{V.}~\bibnamefont{Meden}},
  \bibinfo{author}{\bibfnamefont{S.}~\bibnamefont{Andergassen}},
  \bibinfo{author}{\bibfnamefont{X.}~\bibnamefont{Barnab\'e-Th\'eriault}},
  \bibinfo{author}{\bibfnamefont{W.}~\bibnamefont{Metzner}}, \bibnamefont{and}
  \bibinfo{author}{\bibfnamefont{K.}~\bibnamefont{Sch\"onhammer}},
  \bibinfo{journal}{Phys. Rev. B} \textbf{\bibinfo{volume}{71}},
  \bibinfo{pages}{155401} (\bibinfo{year}{2005}).

\bibitem[{\citenamefont{Barnab\'e-Th\'eriault
  et~al.}(2005{\natexlab{a}})\citenamefont{Barnab\'e-Th\'eriault, Sedeki,
  Meden, and Sch\"onhammer}}]{Barnabe2005}
\bibinfo{author}{\bibfnamefont{X.}~\bibnamefont{Barnab\'e-Th\'eriault}},
  \bibinfo{author}{\bibfnamefont{A.}~\bibnamefont{Sedeki}},
  \bibinfo{author}{\bibfnamefont{V.}~\bibnamefont{Meden}}, \bibnamefont{and}
  \bibinfo{author}{\bibfnamefont{K.}~\bibnamefont{Sch\"onhammer}},
  \bibinfo{journal}{Phys. Rev. Lett.} \textbf{\bibinfo{volume}{94}},
  \bibinfo{pages}{136405} (\bibinfo{year}{2005}{\natexlab{a}}).

\bibitem[{\citenamefont{Barnab\'e-Th\'eriault
  et~al.}(2005{\natexlab{b}})\citenamefont{Barnab\'e-Th\'eriault, Sedeki,
  Meden, and Sch\"onhammer}}]{Barnabe2005a}
\bibinfo{author}{\bibfnamefont{X.}~\bibnamefont{Barnab\'e-Th\'eriault}},
  \bibinfo{author}{\bibfnamefont{A.}~\bibnamefont{Sedeki}},
  \bibinfo{author}{\bibfnamefont{V.}~\bibnamefont{Meden}}, \bibnamefont{and}
  \bibinfo{author}{\bibfnamefont{K.}~\bibnamefont{Sch\"onhammer}},
  \bibinfo{journal}{Phys. Rev. B} \textbf{\bibinfo{volume}{71}},
  \bibinfo{pages}{205327} (\bibinfo{year}{2005}{\natexlab{b}}).

\bibitem[{\citenamefont{Oshikawa et~al.}(2006)\citenamefont{Oshikawa, Chamon,
  and Affleck}}]{Oshikawa2006}
\bibinfo{author}{\bibfnamefont{M.}~\bibnamefont{Oshikawa}},
  \bibinfo{author}{\bibfnamefont{C.}~\bibnamefont{Chamon}}, \bibnamefont{and}
  \bibinfo{author}{\bibfnamefont{I.}~\bibnamefont{Affleck}},
  \bibinfo{journal}{J. Stat. Mech.} \textbf{\bibinfo{volume}{2006}},
  \bibinfo{pages}{P02008} (\bibinfo{year}{2006}).

\bibitem[{\citenamefont{Aristov and W\"{o}lfle}(2009)}]{Aristov2009}
\bibinfo{author}{\bibfnamefont{D.~N.} \bibnamefont{Aristov}} \bibnamefont{and}
  \bibinfo{author}{\bibfnamefont{P.}~\bibnamefont{W\"{o}lfle}},
  \bibinfo{journal}{Phys. Rev. B} \textbf{\bibinfo{volume}{80}},
  \bibinfo{eid}{045109} (\bibinfo{year}{2009}).

\bibitem[{\citenamefont{Yue et~al.}(1994)\citenamefont{Yue, Glazman, and
  Matveev}}]{Yue1994}
\bibinfo{author}{\bibfnamefont{D.}~\bibnamefont{Yue}},
  \bibinfo{author}{\bibfnamefont{L.~I.} \bibnamefont{Glazman}},
  \bibnamefont{and} \bibinfo{author}{\bibfnamefont{K.~A.}
  \bibnamefont{Matveev}}, \bibinfo{journal}{Phys. Rev. B}
  \textbf{\bibinfo{volume}{49}}, \bibinfo{pages}{1966} (\bibinfo{year}{1994}).

\bibitem[{\citenamefont{Das et~al.}(2004)\citenamefont{Das, Rao, and
  Sen}}]{Das2004}
\bibinfo{author}{\bibfnamefont{S.}~\bibnamefont{Das}},
  \bibinfo{author}{\bibfnamefont{S.}~\bibnamefont{Rao}}, \bibnamefont{and}
  \bibinfo{author}{\bibfnamefont{D.}~\bibnamefont{Sen}},
  \bibinfo{journal}{Phys. Rev. B} \textbf{\bibinfo{volume}{70}},
  \bibinfo{pages}{085318} (\bibinfo{year}{2004}).

\bibitem[{\citenamefont{Aristov et~al.}(2010)\citenamefont{Aristov, Dmitriev,
  Gornyi, Kachorovskii, Polyakov, and W\"olfle}}]{Aristov2010}
\bibinfo{author}{\bibfnamefont{D.~N.} \bibnamefont{Aristov}},
  \bibinfo{author}{\bibfnamefont{A.~P.} \bibnamefont{Dmitriev}},
  \bibinfo{author}{\bibfnamefont{I.~V.} \bibnamefont{Gornyi}},
  \bibinfo{author}{\bibfnamefont{V.~Y.} \bibnamefont{Kachorovskii}},
  \bibinfo{author}{\bibfnamefont{D.~G.} \bibnamefont{Polyakov}},
  \bibnamefont{and} \bibinfo{author}{\bibfnamefont{P.}~\bibnamefont{W\"olfle}},
  \bibinfo{journal}{Phys. Rev. Lett.} \textbf{\bibinfo{volume}{105}},
  \bibinfo{pages}{266404} (\bibinfo{year}{2010}).

\bibitem[{\citenamefont{Affleck and Ludwig}(1991)}]{Affleck1991}
\bibinfo{author}{\bibfnamefont{I.}~\bibnamefont{Affleck}} \bibnamefont{and}
  \bibinfo{author}{\bibfnamefont{A.~W.~W.} \bibnamefont{Ludwig}},
  \bibinfo{journal}{Nuclear Physics B} \textbf{\bibinfo{volume}{360}},
  \bibinfo{pages}{641} (\bibinfo{year}{1991}).

\bibitem[{\citenamefont{Fabrizio and Gogolin}(1995)}]{Fabrizio1995}
\bibinfo{author}{\bibfnamefont{M.}~\bibnamefont{Fabrizio}} \bibnamefont{and}
  \bibinfo{author}{\bibfnamefont{A.~O.} \bibnamefont{Gogolin}},
  \bibinfo{journal}{Phys. Rev. B} \textbf{\bibinfo{volume}{51}},
  \bibinfo{pages}{17827} (\bibinfo{year}{1995}).

\bibitem[{\citenamefont{Safi and Schulz}(1995)}]{Safi1995}
\bibinfo{author}{\bibfnamefont{I.}~\bibnamefont{Safi}} \bibnamefont{and}
  \bibinfo{author}{\bibfnamefont{H.~J.} \bibnamefont{Schulz}},
  \bibinfo{journal}{Phys. Rev. B} \textbf{\bibinfo{volume}{52}},
  \bibinfo{pages}{R17040} (\bibinfo{year}{1995}).

\bibitem[{\citenamefont{Aristov}(2011)}]{Aristov2011}
\bibinfo{author}{\bibfnamefont{D.~N.} \bibnamefont{Aristov}},
  \bibinfo{journal}{Phys. Rev. B} \textbf{\bibinfo{volume}{83}},
  \bibinfo{pages}{115446} (\bibinfo{year}{2011}).

\bibitem[{\citenamefont{Bengtsson et~al.}(2005)\citenamefont{Bengtsson,
  Ericsson, Ku{\'s}, Tadej, and {\.Z}yczkowski}}]{Bengtsson2005}
\bibinfo{author}{\bibfnamefont{I.}~\bibnamefont{Bengtsson}},
  \bibinfo{author}{\bibfnamefont{{\AA}.}~\bibnamefont{Ericsson}},
  \bibinfo{author}{\bibfnamefont{M.}~\bibnamefont{Ku{\'s}}},
  \bibinfo{author}{\bibfnamefont{W.}~\bibnamefont{Tadej}}, \bibnamefont{and}
  \bibinfo{author}{\bibfnamefont{K.}~\bibnamefont{{\.Z}yczkowski}},
  \bibinfo{journal}{Commun. Math. Phys.} \textbf{\bibinfo{volume}{259}},
  \bibinfo{pages}{307} (\bibinfo{year}{2005}).

\bibitem[{\citenamefont{Kane and Fisher}(1992)}]{Kane1992}
\bibinfo{author}{\bibfnamefont{C.~L.} \bibnamefont{Kane}} \bibnamefont{and}
  \bibinfo{author}{\bibfnamefont{M.~P.~A.} \bibnamefont{Fisher}},
  \bibinfo{journal}{Phys. Rev. B} \textbf{\bibinfo{volume}{46}},
  \bibinfo{pages}{15233} (\bibinfo{year}{1992}).

\bibitem[{\citenamefont{Aristov}(2007)}]{Aristov2007}
\bibinfo{author}{\bibfnamefont{D.~N.} \bibnamefont{Aristov}},
  \bibinfo{journal}{Phys. Rev. B} \textbf{\bibinfo{volume}{76}},
  \bibinfo{pages}{085327} (\bibinfo{year}{2007}).

\bibitem[{\citenamefont{Tilma and Sudarshan}(2002)}]{Tilma2002}
\bibinfo{author}{\bibfnamefont{T.}~\bibnamefont{Tilma}} \bibnamefont{and}
  \bibinfo{author}{\bibfnamefont{E.~C.~G.} \bibnamefont{Sudarshan}},
  \bibinfo{journal}{Journal of Physics A: Mathematical and General}
  \textbf{\bibinfo{volume}{35}}, \bibinfo{pages}{10467} (\bibinfo{year}{2002}).
  
\bibitem{misprint} The definition of $b_{4}$ after Eq.\ (49) in [\onlinecite{Aristov2009}]
has a  wrong sign, it should read $(2\ln2- 2)$ and $(4\ln2 -2)$ for  $T=0$ and 
$T\neq 0$, respectively.


\end{thebibliography}

\end{document}